\documentclass[a4paper,11pt]{article}

\usepackage{jcappub} 

\usepackage[T1]{fontenc} 
\usepackage{graphicx}
\graphicspath{{figures/}}
\usepackage[utf8]{inputenc}
\usepackage{amsmath}
\usepackage{amsfonts}
\usepackage{amssymb}
\usepackage{enumitem}
\usepackage{multirow}
\usepackage{orcidlink}
\usepackage{CJKutf8}
\usepackage{color}
\usepackage[capitalise]{cleveref}
\usepackage{acro}
\usepackage{adjustbox}
\usepackage{array}
\usepackage{natbib}
\usepackage{dblfnote}
\DFNalwaysdouble
\usepackage{slashed}
\usepackage{dcolumn}
\usepackage{hyperref}
\usepackage{geometry}
\usepackage{subfig}
\usepackage{overpic}
\usepackage{inputenc}
\usepackage{caption}
\usepackage[theorems,skins]{tcolorbox}
\usepackage{booktabs}
\usepackage{tabularx}
\usepackage{array}
\usepackage{ragged2e}

\newcolumntype{L}[1]{>{\RaggedRight\arraybackslash}p{#1}}
\newcolumntype{Y}{>{\RaggedRight\arraybackslash}X}

\def\({\left(}
\def\){\right)}
\def\[{\left[}
\def\]{\right]}
\def\be{\begin{eqnarray}}
\def\ee{\end{eqnarray}}

\DeclareAcronym{GW}{
  short = GW ,
  long = gravitational wave ,
  short-plural = s
}
\DeclareAcronym{LIGO}{
  short = LIGO ,
  long = Laser Interferometer Gravitational-wave Observatory ,
  short-plural =
}
\DeclareAcronym{LISA}{
  short = LISA ,
  long = Laser Interferometer Space Antenna ,
  short-plural =
}
\DeclareAcronym{SKA}{
  short = SKA ,
  long = Square Kilometre Array ,
  short-plural =
}

\DeclareAcronym{SNR}{
	short = SNR ,
	long = signal-to-noise ratio ,
	short-plural =
}

\DeclareAcronym{PTA}{
	short = PTA ,
	long = pulsar timing array ,
	short-plural =
}

\DeclareAcronym{FLRW}{
  short = FLRW ,
  long = Friedmann-Lemaitre-Robertson-Walker ,
  short-plural =
}

\DeclareAcronym{SIGW}{
	short = SIGW ,
	long = scalar induced gravitational wave ,
	short-plural =  s
}

\DeclareAcronym{PBH}{
	short = PBH ,
	long = primordial black hole ,
	short-plural =  s
}

\DeclareAcronym{SMBHB}{
  short = SMBHB ,
  long = supermassive black hole binary ,
  short-plural = s
}

\DeclareAcronym{1PI}{
	short = 1PI ,
	long = one-particle irreducible  ,
	short-plural =
}

\DeclareAcronym{1PR}{
	short = 1PR ,
	long = one-particle reducible  ,
	short-plural =
}

\DeclareAcronym{KDE}{
  short = KDE ,
  long = kernel density estimator ,
  short-plural = s
}

\DeclareAcronym{BPBHM}{
  short = BPBHM ,
  long = binary primordial black hole merger,
  short-plural = s
}

\DeclareAcronym{CMB}{
	short = CMB ,
	long = cosmic microwave background ,
	short-plural =
}
\DeclareAcronym{DM}{
	short = DM ,
	long = dark matter ,
	short-plural =
}

\DeclareAcronym{BBN}{
	short = BBN ,
	long = Big-Bang nucleosynthesis ,
	short-plural =
}

\DeclareAcronym{LN}{
	short = LN ,
	long = log-normal  ,
	short-plural =
}

\DeclareAcronym{BPL}{
	short = BPL ,
	long = broken power-law ,
	short-plural =
}

\DeclareAcronym{SGWB}{
	short = SGWB ,
	long = stochastic gravitational	wave background ,
	short-plural =  s
}

\DeclareAcronym{LSS}{
	short = LSS ,
	long = large scale structure ,
	short-plural =
}

\DeclareAcronym{RD}{
	short = RD ,
	long = radiation-dominated ,
	short-plural =
}

\DeclareAcronym{PLS}{
	short = PLS ,
	long = power low sensitivity ,
	short-plural =
}

\DeclareAcronym{MAP}{
	short = MAP ,
	long = maximum a posterior ,
	short-plural =
}

\DeclareAcronym{PGW}{
	short = PGW ,
	long = primordial gravitational wave ,
	short-plural =  s
}

\DeclareAcronym{BAO}{
	short = BAO ,
	long = baryon acoustic oscillations ,
	short-plural =
}

\DeclareAcronym{TSIGW}{
	short = TSIGW ,
	long = tensor-scalar induced gravitational wave ,
	short-plural =  s
}

\title{\boldmath Angular power spectrum of  induced gravitational waves: breaking model and parameter degeneracies in PTA observations}

\author[a]{Peng-Yu Wu,}
\author[b]{Di Wu,}
\author[c,1]{Jing-Zhi Zhou,\note{Corresponding author.}  }

\affiliation[a]{Center for Joint Quantum Studies and Department of Physics,
School of Science, Tianjin University, Tianjin 300350, China}
\affiliation[b]{School of Fundamental Physics and Mathematical Sciences, Hangzhou Institute for Advanced Study, University of Chinese Academy of Sciences, Hangzhou 310024, China}

\affiliation[c]{ Department of Mathematics and Physics, Huai'an University, Meicheng East Road 1, Huai'an, Jiangsu 223200, China}

\emailAdd{wupy2023@tju.edu.cn}
\emailAdd{wudi@ucas.ac.cn}
\emailAdd{zhoujingzhi@hau.edu.cn}

\abstract{ Large primordial curvature and tensor perturbations on small scales can generate second-order tensor-scalar induced gravitational waves (TSIGWs) through scalar-scalar, tensor-scalar, and tensor-tensor source terms. TSIGWs are recognized as one of the principal contributors to the stochastic gravitational wave background (SGWB) observations. We explore the energy density spectrum of second-order TSIGWs in the presence of non-Gaussian primordial curvature perturbations. Furthermore, we present the first calculation of the anisotropic angular power spectrum associated with second-order TSIGWs, which provides a promising tool to discriminate them from other SGWB sources. }

\begin{document}

\maketitle
\flushbottom

\section{Introduction}\label{sec:1.0}
With the detection of \acp{GW} from black hole and neutron star mergers by LIGO/Virgo, gravitational wave astronomy has entered a new era \cite{LIGOScientific:2016aoc,LIGOScientific:2017vwq,LIGOScientific:2016sjg,LIGOScientific:2017ycc}. These pioneering observations have raised expectations that studies of the \ac{SGWB} may reveal essential clues about the early Universe. In June 2023, the \ac{PTA} collaborations NANOGrav \cite{NANOGrav:2023gor,NANOGrav:2023hde}, EPTA \cite{EPTA:2023fyk}, PPTA \cite{Reardon:2023gzh}, and CPTA \cite{Xu:2023wog} reported evidence for an isotropic, stochastic background of \acp{GW} in the nHz frequency range. This milestone result has opened a fresh observational window onto new physics in the evolution of the early Universe and the characterization of primordial perturbations on small scales.

The \ac{SGWB} detected in the nHz frequency band has an origin that remains ambiguous. Multiple candidate sources have been proposed. From an astrophysical perspective, the PTA signal is most consistently explained by \ac{SMBHB} \cite{NANOGrav:2023hfp,Middleton:2020asl,NANOGrav:2020spf,Bi:2023tib}. However, cosmological scenarios provide equally compelling alternatives, such as phase transitions \cite{Athron:2023xlk,Fujikura:2023lkn,Addazi:2023jvg,Jiang:2023qbm,Wu:2023hsa,He:2023ado},  cosmic strings \cite{Ellis:2020ena,Lazarides:2023ksx,Yamada:2023thl,Ellis:2023tsl}, and \acp{SIGW} \cite{Domenech:2026nun,Inui:2024fgk,Wang:2023div,Pi:2024lsu,Vaskonen:2020lbd,DeLuca:2020agl,Balaji:2023ehk,Franciolini:2023pbf,Zhu:2023gmx,Chang:2023vjk,Lozanov:2023rcd,Yu:2025cqu,Liu:2023ymk,Liu:2023pau,Zhao:2026jax,Zhao:2022kvz,Cai:2023dls,Xie:2026xhr,Papanikolaou:2024fzf,Zhang:2026hqg,Caravano:2026hca,Perna:2026szd,Domenech:2025ccu,Bhaumik:2026rzf,Chen:2026scy,Marriott-Best:2026ewq,Huang:2026bwa}. Since the dominant source of the \ac{PTA} signal cannot yet be determined with certainty, current studies typically assume that the \ac{SGWB} generated by a specific mechanism dominates the current observations. For each proposed mechanism, theoretical computations provide the associated energy density spectrum, which is sensitive to one or more parameters. Variations in these parameters can significantly affect the spectral features, including its frequency range, amplitude, and overall shape. By comparing the \ac{PTA} data on the energy density spectrum with theoretical predictions, one can perform rigorous Bayesian analyses to constrain the parameter space of different candidate models for the dominant \ac{PTA} source. Furthermore, the calculation of Bayes factors for various candidate models provides a systematic and quantitative framework to examine the plausibility that different \ac{SGWB} origins dominate the current \ac{PTA} observations \cite{Zhou:2025djn,Wang:2025kbj,NANOGrav:2023hvm,You:2023rmn,Sesana:2025udx}. In summary, current analyses of the \ac{PTA} signal typically focus on two main aspects: (1) computing the energy density spectrum under a given model and constraining its parameter space using \ac{PTA} data; and (2) comparing the Bayes factors of different models to evaluate their feasibility as the dominant source of the current \ac{PTA} observations. However, due to the limited precision of current PTA measurements, it remains difficult to completely exclude certain models as potential dominant contributors based solely on these two approaches. The ability to distinguish between different \ac{SGWB} sources and to establish the dominant origin of \ac{PTA}  observations represents one of the most important directions for future \ac{SGWB} research.

To investigate the dominant source of the \ac{SGWB} in the nHz frequency band, relying solely on \ac{PTA} observations of the corresponding energy density spectrum is far from sufficient. As noted earlier, for many \ac{SGWB} generation mechanisms, even with more precise future measurements of the energy density spectrum, theoretical predictions can always be adjusted to fit the data by tuning model parameters or modifying assumptions. To resolve this challenge, constraints from other observations must be considered. Current studies commonly adopt three approaches:

\noindent
\textbf{(1) Other cosmological observations:} considering indirect observational bounds from experimental data such as \ac{CMB}, \ac{BAO}, $N_\mathrm{eff}$, and \acp{PBH}, which can effectively reduce the allowed parameter space of certain models and thereby better constrain or exclude them \cite{Cang:2023ysz,Zhou:2024yke,Ben-Dayan:2025bqd,Wang:2025qpj,Wang:2023sij,Ben-Dayan:2019gll,Bugaev:2010bb}.

\noindent
\textbf{(2) \ac{SGWB} in other frequency bands:} examining future \ac{SGWB} observations in other frequency bands, such as those expected from \ac{LISA}, to test whether a model dominating the \ac{PTA} signal would also produce detectable signatures in the \ac{LISA} band \cite{LISACosmologyWorkingGroup:2025vdz,LISACosmologyWorkingGroup:2024hsc,LISACosmologyWorkingGroup:2023njw,LISACosmologyWorkingGroup:2022jok,Maiti:2026dsl}.

\noindent
\textbf{(3) Anisotropy in \ac{SGWB}:} analyzing differences in the anisotropic angular power spectra predicted by various \ac{SGWB} generation mechanisms, which, combined with future high-precision \ac{SGWB} measurements, will provide a crucial window for distinguishing among different sources \cite{LISACosmologyWorkingGroup:2022kbp,NANOGrav:2023tcn,Wang:2023ost,Ruiz:2024weh,Depta:2024ykq,Lemke:2024cdu,Sah:2024oyg,Zhao:2024yau,Ding:2023xeg,Contaldi:2016koz,Bellomo:2021mer,ValbusaDallArmi:2020ifo,Malhotra:2022ply,Dimastrogiovanni:2021mfs,Wang:2021djr,Liu:2020mru,Chung:2023rpq,Zhao:2024gan}.

\noindent
In summary, the primary roles of considering observations beyond the energy density spectrum of the \ac{SGWB} in the nHz band are twofold: first, to better constrain the parameter space allowed by current measurements; and second, to incorporate observations of other types or frequency bands of \acp{SGWB} in order to more effectively distinguish their different origins. By employing the aforementioned methods together with future, more precise cosmological observations, we may be able to better identify the sources of various \acp{SGWB} in the near future.

As discussed earlier, we can rigorously quantify the possibility that different models dominate the current \ac{PTA} observations by directly computing the Bayes factors between them. In Ref.~\cite{NANOGrav:2023hvm}, the NANOGrav collaboration investigated various models of the \ac{SGWB} in the nHz band and calculated the corresponding Bayes factors. The results indicate that second-order \acp{SIGW} yield the highest Bayes factor, suggesting that \acp{SIGW} might be more likely to dominate the current \ac{PTA} observations. \acp{SIGW} originate from large-amplitude primordial curvature perturbations on small scales. Specifically, when deriving the second-order cosmological perturbation equations, the equation of motion of second-order tensor perturbations contains numerous terms composed of products of two first-order scalar perturbations. Consequently, if the primordial scalar perturbations are sufficiently large, the first-order scalar modes can couple to the second-order tensor perturbations through the second-order cosmological perturbation equations, thereby generating the \ac{SGWB} with significant observable effects. It should be noted that on large scales ($\gtrsim$1 Mpc), the power spectrum of primordial curvature perturbations is approximately scale-invariant, with an amplitude of $A_{\zeta} \approx 2\times 10^{-9}$  \cite{Planck:2018vyg}. However, unlike large-scale primordial perturbations, there are no stringent observational constraints on small-scale ($\lesssim$1 Mpc) primordial perturbations \cite{Bringmann:2011ut}. Therefore, large-amplitude primordial curvature perturbations can only exist at small scales, potentially generating \acp{SIGW} that are directly detectable. Similar to primordial curvature perturbations, current cosmological observations do not impose strict constraints on primordial tensor perturbations at small scales. When large primordial curvature and tensor perturbations coexist on small scales, the equation of motion of second-order tensor perturbations contains three types of source terms formed by products of first-order cosmological perturbations: scalar-scalar, tensor-scalar, and tensor-tensor. The second-order gravitational waves jointly induced by primordial curvature and tensor perturbations are referred to as second-order \acp{TSIGW} \cite{Chang:2022vlv,Picard:2024ekd,Picard:2023sbz,Bari:2023rcw,Picard:2025bwq,Yu:2023lmo,Iania:2026mtp}. Analogous to \acp{SIGW}, \acp{TSIGW} may also dominate current \ac{PTA} observations \cite{Wu:2025gwt}.

The study of \acp{SIGW} is inspired by prior analyses of second-order tensor and vector perturbations induced by first-order scalars and their impact on CMB polarization \cite{Mollerach:2003nq}. Building on this foundation, Refs.~\cite{Ananda:2006af,Osano:2006ew,Baumann:2007zm} examined second-order gravitational waves arising from large-amplitude primordial scalar perturbations at small scales and calculated the associated energy density spectrum. In the past decade, research on second-order \acp{SIGW} has been extended in various directions \cite{Domenech:2021ztg}. Moreover, in Refs.~\cite{Bartolo:2019oiq,Bartolo:2019yeu,Li:2023qua,Li:2023xtl,Li:2025met,Dimastrogiovanni:2022eir}, the anisotropic angular power spectrum of second-order \acp{SIGW} has been systematically investigated. In this paper, we will calculate the energy density spectrum and the corresponding angular power spectrum of second-order \acp{TSIGW} in the presence of local-type non-Gaussian primordial curvature perturbations. We demonstrate that parameter degeneracies, or the observational indistinguishability of energy density spectra from different models, make the anisotropic angular power spectrum a powerful diagnostic tool for identifying the origin of the \ac{SGWB}.

This paper is organized as follows. In Sec.~\ref{sec:2.0}, we review the theoretical results of \acp{SIGW} and \acp{TSIGW}. In Sec.~\ref{sec:3.0}, we discuss the energy density spectra of \acp{SIGW} and \acp{TSIGW} in the presence of local-type primordial non‑Gaussianity. In Sec.~\ref{sec:4}, we investigate the angular power spectrum of \acp{TSIGW}. In Sec.~\ref{sec:Obs}, we apply the theoretical results obtained in the previous sections to \ac{SGWB} observations. Finally, we summarize our results and give some discussion in Sec.~\ref{sec:6}.

\section{Second-order TSIGWs}\label{sec:2.0}
In this section, we review the key results for \acp{TSIGW} sourced by Gaussian primordial curvature perturbations and primordial tensor perturbations. In the Newtonian gauge, the perturbed cosmological metric takes the following form:
\begin{eqnarray}\label{eq:dS}
	\mathrm{d} s^{2}=a^{2}\left[-\left(1+2 \phi^{(1)}\right) \mathrm{d} \eta^{2}+\left(\left(1-2 \psi^{(1)}\right) \delta_{i j}+h^{(1)}_{ij}+\frac{1}{2}h^{(2)}_{ij}\right)\mathrm{d} x^{i} \mathrm{d} x^{j}\right] \ ,
\end{eqnarray}
where $\phi^{(1)}$ and $\psi^{(1)}$ are first-order scalar perturbations, and $h^{(n)}_{ij}$ $\left( n=1,2 \right)$ are $n$th-order tensor perturbations. The equation of motion for the second-order \acp{TSIGW} during the \ac{RD} era is given by
\begin{equation}\label{eq:h}
	\begin{aligned}
		h_{ij}^{(2)''}(\eta,\mathbf{x})+2 \mathcal{H} h_{ij}^{(2)'}(\eta,\mathbf{x})-\Delta h_{ij}^{(2)}(\eta,\mathbf{x})=-4 \Lambda_{ij}^{lm} \left( \mathcal{S}^{(2)}_{lm,\phi\phi}+\mathcal{S}^{(2)}_{lm,\phi h}+\mathcal{S}^{(2)}_{lm,hh}\right)   \ .
	\end{aligned}
\end{equation}
Eq.~(\ref{eq:h}) can be obtained by computing the second-order cosmological perturbation equations and extracting their transverse-traceless part. The explicit expressions of the source terms $\mathcal{S}^{(2)}_{lm,\phi\phi}$, $\mathcal{S}^{(2)}_{lm,\phi h}$, and $\mathcal{S}^{(2)}_{lm,hh}$ are given as follows
\begin{eqnarray}
	\mathcal{S}^{(2)}_{lm,\phi\phi}(\eta,\mathbf{x})&=&\partial_{l} \phi^{(1)} \partial_{m} \phi^{(1)} +4 \phi^{(1)} \partial_{l} \partial_{m} \phi^{(1)}-\frac{1}{ \mathcal{H}}\left(\partial_{l} \phi^{(1)'} \partial_{m} \phi^{(1)}+\partial_{l} \phi^{(1)}  \partial_{m} \phi^{(1)'}\right) \nonumber\\
	&-&\frac{1}{ \mathcal{H}^{2}} \partial_{l} \phi^{(1)'} \partial_{m}  \phi^{(1)'} \ ,
	\label{eq:1}
    \end{eqnarray}
\begin{eqnarray}
	\mathcal{S}^{(2)}_{lm,\phi h}(\eta,\mathbf{x})&=& 10\mathcal{H}h^{(1)}_{lm}\phi^{(1)'}+3h^{(1)}_{lm}\phi^{(1)''}-2\partial_{b}h^{(1)}_{lm}\partial^{b}\phi^{(1)}-2\phi^{(1)}\Delta h^{(1)}_{lm} -\frac{5}{3}h^{(1)}_{lm}\Delta\phi^{(1)} \ , \label{eq:2}
\end{eqnarray}
\begin{eqnarray}
	\mathcal{S}^{(2)}_{lm,hh}(\eta,\mathbf{x})&=&\frac{1}{2}\left( -h^{b,(1)'}_{l} h^{(1)'}_{mb} +\partial_c h^{(1)}_{mb}\partial^c h^{b,(1)}_{l} -h^{bc,(1)}\partial_c \partial_m h^{(1)}_{lb}  -\partial_b h^{(1)}_{mc}\partial^c h^{b,(1)}_{l}  \right. \nonumber\\
	&+&\left.\frac{1}{2} h^{bc,(1)}\partial_l \partial_m h^{(1)}_{bc}+ h^{bc,(1)}\partial_c \partial_b h^{(1)}_{lm}-h^{bc,(1)}\partial_c \partial_l h^{(1)}_{mb} \right) \ .
	\label{eq:3}
\end{eqnarray}
During the \ac{RD} era, we set $\mathcal{H}=a'/a=1/\eta$ and $w=c_s^2=1/3$. In Eq.~(\ref{eq:1})--Eq.~(\ref{eq:3}), the three source terms are associated with the first-order scalar-scalar contribution, the scalar-tensor contribution, and the tensor-tensor contribution, respectively. When considering large-amplitude primordial curvature and tensor perturbations on small scales, the source terms in the equation of motion for second-order induced \acp{GW} involve first-order scalar and tensor perturbations. Accordingly, the analysis requires solving the first-order scalar and tensor perturbations in advance. The equations of motion for these two types of first-order cosmological perturbations during the \ac{RD} era are given by
\begin{equation}\label{eq:h100}
	\begin{aligned}
		h_{ij}^{(1)''}(\eta,\mathbf{x})+2 \mathcal{H} h_{ij}^{(1)'}(\eta,\mathbf{x})-\Delta h_{ij}^{(1)}(\eta,\mathbf{x})=0 \ ,
	\end{aligned}
\end{equation}
\begin{eqnarray}
  &&\psi^{(1)}-\phi^{(1)}=0 \ , \label{eq:s11}
      \\ \nonumber\\
&&2\mathcal{H} \left(\phi^{(1)'}+3\psi^{(1)'}   \right)  +2\psi^{(1)''}+\Delta \phi^{(1)}  -\Delta \psi^{(1)}-\frac{2}{3}\Delta \psi^{(1)}=0 \ . \label{eq:s12}
\end{eqnarray}
In momentum space, the two types of first-order cosmological perturbations described above can be expressed in terms of the primordial perturbations and the first-order transfer functions as follows
\begin{eqnarray}
   &&\phi^{(1)}_{\mathbf{q}}(\eta)=\psi^{(1)}_{\mathbf{q}}(\eta)=\frac{2}{3}\zeta_{\mathbf{q}}T_\phi(q\eta)  \ , \label{eq:Ts1} \\
   &&h^{\lambda,(1)}_{\mathbf{q}}(\eta)=h^{\lambda}_{\mathbf{q}}T_h(q\eta) \label{eq:Tt1} \ ,
\end{eqnarray}
where $h^{\lambda,(n)}_{\mathbf{q}}(\eta)=\varepsilon^{\lambda,ij}(\mathbf{q})h^{(n)}_{ij}(\eta,\mathbf{q})$, and $\varepsilon^{\lambda,ij}(\mathbf{q})$ denotes the polarization tensor associated with the momentum $\mathbf{q}$. The symbol $q\equiv|\mathbf{q}|$ denotes the magnitude of the vector $\mathbf{q}$. $T_\phi(q\eta)$ and $T_h(q\eta)$ represent the transfer functions of the first-order scalar and tensor perturbations, respectively. $\zeta_{\mathbf{q}}$ and $h^{\lambda}_{\mathbf{q}}$ denote the large-amplitude primordial curvature and tensor perturbations generated during inflation. By substituting Eq.~(\ref{eq:Ts1}) and Eq.~(\ref{eq:Tt1}) into Eq.~(\ref{eq:h100})--Eq.~(\ref{eq:s12}), we obtain the equations of motion of two first-order transfer functions
\begin{equation}\label{eq:eTs}
\begin{aligned}
    \frac{\mathrm{d}^2}{\mathrm{d}x^2}T_{\phi}(x)&+\frac{4}{x} \frac{\mathrm{d}}{\mathrm{d}x}T_{\phi}(x) +\frac{1}{3} T_{\phi}(x) = 0 \ ,
    \end{aligned}
\end{equation}
\begin{equation}\label{eq:eTt}
\begin{aligned}
    \frac{\mathrm{d}^2}{\mathrm{d}x^2}T_{h}(x)&+\frac{2}{x} \frac{\mathrm{d}}{\mathrm{d}x}T_{h}(x) + T_{h}(x) = 0 \ .
    \end{aligned}
\end{equation}
where $x=q\eta$. The analytical solutions of the two transfer functions in Eq.~(\ref{eq:eTs}) and Eq.~(\ref{eq:eTt}) are given by
\begin{eqnarray}\label{eq:Tas}
    T_{\phi}(x)=\frac{9}{x^{2}}\left(\frac{\sqrt{3}}{x} \sin \left(\frac{x}{\sqrt{3}}\right)-\cos \left(\frac{x}{\sqrt{3}}\right)\right) \ , \
    T_{h}(x)=\frac{\sin{x}}{x} \ .
\end{eqnarray}
After obtaining solutions for both the first-order scalar and tensor perturbations, the source term in the equation of motion of the second-order \acp{TSIGW} in Eq.~(\ref{eq:h}) becomes a fully determined analytic function. Using the Green's function approach to solve Eq.~(\ref{eq:h}), we obtain the expression for second-order \acp{TSIGW} as
\begin{equation}\label{eq:hsum}
	\begin{aligned}
		h^{\lambda,(2)}(\eta,\mathbf{q})=\sum^{7}_{i=1} h_i^{\lambda,(2)}(\eta,\mathbf{q}) \ ,
	\end{aligned}
\end{equation}
where
\begin{eqnarray}
	h_1^{\lambda,(2)}(\eta,\mathbf{q})&=&\frac{4}{9}\int\frac{\mathrm{d}^3k}{(2\pi)^{3/2}}\varepsilon^{\lambda,lm}\left(\mathbf{q}\right)k_lk_mI^{(2)}_{1}\left( u,v,x \right)\zeta_{\mathbf{q}-\mathbf{k}}\zeta_{\mathbf{k}} \ ,
	\label{eq:hf1}\\
	h_2^{\lambda,(2)}(\eta,\mathbf{q})&=&\frac{2}{3}\int\frac{\mathrm{d}^3k}{(2\pi)^{3/2}}\varepsilon^{\lambda,lm}\left(\mathbf{q}\right) \varepsilon^{\lambda_1}_{lm}\left(\mathbf{k}\right)q^2 I^{(2)}_{2}\left( u,v,x \right) \zeta_{\mathbf{q}-\mathbf{k}}\mathbf{h}^{\lambda_1}_{\mathbf{k}} \ ,
	\label{eq:hf2}\\
	h_i^{\lambda,(2)}(\eta,\mathbf{q})&=&\int\frac{\mathrm{d}^3k}{(2\pi)^{3/2}}\varepsilon^{\lambda,lm}\left(\mathbf{q}\right) \mathbb{P}_{lm,i}^{\lambda_1\lambda_2}I^{(2)}_{i}\left(u,v,x\right)\mathbf{h}^{\lambda_1}_{\mathbf{q}-\mathbf{k}}\mathbf{h}^{\lambda_2}_{\mathbf{k}} \ , \ (i=3\sim7) \ . \label{eq:hfi}
\end{eqnarray}
In Eq.~(\ref{eq:hf1})--Eq.~(\ref{eq:hfi}), the three formal expressions correspond respectively to the second-order \acp{TSIGW} generated by three different source terms. These expressions can be systematically decomposed into four components: momentum integration: $\int\mathrm{d}^3k$; momentum polynomials: $\varepsilon^{\lambda,lm}\left(\mathbf{q}\right)k_lk_m$, $\varepsilon^{\lambda,lm}\left(\mathbf{q}\right) \varepsilon^{\lambda_1}_{lm}\left(\mathbf{k}\right)q^2$, and $\mathbb{P}_{lm,i}^{\lambda_1\lambda_2}$; primordial perturbations: $\zeta_{\mathbf{q}}$ and $\mathbf{h}^{\lambda_1}_{\mathbf{q}}$; and kernel functions: $I^{(2)}_i$ $(i=1\sim7)$. More precisely, the momentum integration originates from the convolution formula applied in the Fourier transformation of the source terms in Eq.~(\ref{eq:1})--Eq.~(\ref{eq:3}), with its multiplicity increasing when higher-order induced \acp{GW} or primordial non-Gaussianities are taken into account. The momentum polynomials in Eq.~(\ref{eq:hf1})--Eq.~(\ref{eq:hfi}) arise from partial derivatives in the source terms of second-order \acp{TSIGW} and the polarization tensor of the first-order tensor perturbations. The explicit expression of the momentum polynomial $\mathbb{P}_{lm,i}^{\lambda_1\lambda_2}$ in Eq.~(\ref{eq:hfi}) is given by
\begin{equation}
\begin{aligned}
\mathbb{P}_{lm,3}^{\lambda_1\lambda_2}&=\varepsilon^{\lambda,lm}\left(\mathbf{q}\right) \varepsilon^{\lambda_1,b}_{l}\left(\mathbf{q}-\mathbf{k}\right)\varepsilon^{\lambda_2}_{bm}\left(\mathbf{k}\right)q^2 \ , \ \mathbb{P}_{lm,4}^{\lambda_1\lambda_2}=2\varepsilon^{\lambda,lm}\left(\mathbf{q}\right) \varepsilon^{\lambda_1,bc}\left( \mathbf{q}-\mathbf{k} \right)\varepsilon^{\lambda_2}_{mb}\left( \mathbf{k} \right)k_ck_l \ , \\
\mathbb{P}_{lm,5}^{\lambda_1\lambda_2}&=\varepsilon^{\lambda,lm}\left(\mathbf{q}\right) \varepsilon^{\lambda_1}_{mc}\left(\mathbf{q}-\mathbf{k}\right)\varepsilon^{\lambda_2}_{lb}\left(\mathbf{k}\right)\left( q-k \right)^bk^c \ , \ \mathbb{P}_{lm,6}^{\lambda_1\lambda_2}=\varepsilon^{\lambda,lm}\left(\mathbf{q}\right) \varepsilon^{\lambda_1}_{bc}\left(\mathbf{q}-\mathbf{k}\right)\varepsilon^{\lambda_2}_{lm}\left(\mathbf{k}\right)k^bk^c \ , \\
\mathbb{P}_{lm,7}^{\lambda_1\lambda_2}&=\frac{1}{2}\varepsilon^{\lambda,lm}\left(\mathbf{q}\right) \varepsilon^{\lambda_1,bc}\left(\mathbf{q}-\mathbf{k}\right)\varepsilon^{\lambda_2}_{bc}\left(\mathbf{k}\right)k_lk_m \ .
	\end{aligned}
\end{equation}
Furthermore, the primordial perturbations $\zeta_{\mathbf{q}}$ and $\mathbf{h}^{\lambda_1}_{\mathbf{q}}$ originate from the relations between first-order cosmological perturbations and transfer functions given in Eq.~(\ref{eq:Ts1}) and Eq.~(\ref{eq:Tt1}). Finally, the kernel function $I^{(2)}_i$ $(i=1\sim7)$ describes the dynamical evolution of second-order \acp{TSIGW} in the \ac{RD} era, and by employing the Green’s function method, it can be formulated as an integral over $\bar{x}$
\begin{equation}\label{eq:I}
	\begin{aligned}
		I^{(2)}_i\left( u,v,x \right)=\frac{4}{q^2} \int_{0}^{x} \mathrm{d} \bar{x} \left( \frac{\bar{x}}{x}\sin\left( x-\bar{x} \right) f_i\left( u,v,\bar{x} \right)  \right) \ , \ (i=1 \sim 7) \ ,
	\end{aligned}
\end{equation}
where $k=vq$ and $|\mathbf{q}-\mathbf{k}|=uq$. The functions $f_i^{(2)}$$(i=1\sim7)$ in Eq.~(\ref{eq:f1})--Eq.~(\ref{eq:f4}) can be obtained by substituting the first-order transfer functions into the source terms of second-order \acp{TSIGW} and isolating the scalar components. The explicit expressions of $f_i^{(2)}$$(i=1\sim7)$ are given by
\begin{eqnarray}
	f_1^{(2)}\left( u,v,x \right)&=&x^2uv\frac{\mathrm{d}}{\mathrm{d}(ux)} T_{\phi}\left( ux \right)\frac{\mathrm{d}}{\mathrm{d}(vx)}T_{\phi}\left( vx \right)+2ux T_{\phi}\left(vx \right)\frac{\mathrm{d}}{\mathrm{d}(ux)}T_{\phi}\left( ux \right)\nonumber\\
 &+&3T_{\phi}\left(ux \right)T_{\phi}\left( vx \right) \, ,
	\label{eq:f1}\\
	f_2^{(2)}\left( u,v,x \right)&=&\frac{10u}{x}T_{h}\left( vx \right)\frac{\mathrm{d}}{\mathrm{d}(ux)} T_{\phi}\left( ux \right)+3u^2T_{h}\left( vx \right)\frac{\mathrm{d}^2}{\mathrm{d}(ux)^2}T_{\phi}\left(ux \right)+\frac{5}{3}u^2 T_{\phi}\left( ux \right)T_{h}\left( vx \right) \nonumber\\
	&+&\left(1-v^2-u^2\right)T_{\phi}\left( ux \right)T_{h}\left( vx \right)+2v^2T_{\phi}\left( ux \right)T_{h}\left( vx \right) \, ,
	\label{eq:f2}\\
	f_{3}^{(2)}\left(u,v,x\right)&=&-\frac{1-u^2-v^2}{4}T_{h}\left( ux \right)T_{h}\left( vx \right)-\frac{uv}{2}\frac{\mathrm{d}}{\mathrm{d}(ux)}T_{h}\left( ux \right)\frac{\mathrm{d}}{\mathrm{d}(vx)}T_{h}\left( vx \right) \, ,
	\label{eq:f3} \\
	f_i^{(2)}\left( u,v,x \right)&=&\frac{1}{2}T_{h}\left( ux \right)T_{h}\left( vx \right) \ ,  \  (i=4,5,6,7)
	\label{eq:f4} \ .
\end{eqnarray}
By computing the integral in Eq.~(\ref{eq:I}) and simplifying, we obtain the analytic expressions of the kernel functions
\begin{eqnarray}
        I^{(2)}_1(u,v,x)&=&
        \frac{27 (u^2 + v^2 - 3)}{4 q^2u^3 v^3 x}
        \Bigg\{ \cos{x}(u^2 + v^2 - 3)
        \bigg[\mathrm{Si}\left(\left(1-\frac{u+v}{\sqrt{3}}\right) x \right)
        +\mathrm{Si}\left(\left(1+\frac{u+v}{\sqrt{3}}\right) x \right)\nonumber\\
        &-&\mathrm{Si}\left(\left(1+\frac{u-v}{\sqrt{3}}\right) x \right)
        -\mathrm{Si}\left(\left(1-\frac{u-v}{\sqrt{3}}\right) x \right)
        \bigg]
        -\sin{x}\bigg\{4 u v-(u^2 + v^2 - 3)\nonumber\\
        &\times& \bigg[\mathrm{Ci}\left(\left(1+\frac{u-v}{\sqrt{3}}\right) x \right)
        +\mathrm{Ci}\left(\left(1-\frac{u-v}{\sqrt{3}}\right) x \right)
        -\mathrm{Ci}\left(\left(1+\frac{u+v}{\sqrt{3}}\right) x \right)\nonumber\\
        &-&\mathrm{Ci}\left(\left(\bigg|1-\frac{u+v}{\sqrt{3}}\bigg|\right) x \right)
        +\ln \left(\bigg|\frac{3-(u+v)^2}{3-(u-v)^2} \bigg| \right)
        \bigg]
        \bigg\}
            \Bigg\}+\mathcal{O}(\frac{1}{x^2})  \ ,
\end{eqnarray}
\begin{eqnarray}
        I^{(2)}_2(u,v,x) &=&  \frac{\sqrt{3}\cos x}{4 q^2 u^3 v x} \left(u^2 - 3(v-1)^2\right) \left(u^2 - 3(v+1)^2\right)  \nonumber \\
        &\times& \Big[ \mathrm{Si}\left(\left(1-\frac{u}{\sqrt{3}}+v\right)x\right) - \mathrm{Si}\left(\left(1+\frac{u}{\sqrt{3}}+v\right)x\right) \nonumber \\
        &-& \mathrm{Si}\left(\left(1-\frac{u}{\sqrt{3}}-v\right)x\right) + \mathrm{Si}\left(\left(1+\frac{u}{\sqrt{3}}-v\right)x\right) \Big] \nonumber \\
        &-&\frac{\sin x}{4 q^2 u^3 v x} \Bigg\{ 4 u v (9+u^2-9v^2)+\sqrt{3} \left(u^2 - 3(v-1)^2\right) \left(u^2 - 3(v+1)^2\right)   \nonumber \\
        &\times& \Big[ \mathrm{Ci}\left(\left(1-\frac{u}{\sqrt{3}}+v\right)x\right) - \mathrm{Ci}\left(\left(1+\frac{u}{\sqrt{3}}+v\right)x\right) \nonumber \\
        &-& \mathrm{Ci}\left( \left|1-\frac{u}{\sqrt{3}}-v\right| x \right) + \mathrm{Ci}\left( \left|1+\frac{u}{\sqrt{3}}-v\right| x \right)  \nonumber +  \ln\left| \frac{3-(u+\sqrt{3}v)^2}{3-(u-\sqrt{3}v)^2} \right| \Big] \Bigg\} \nonumber \\
    &+& \mathcal{O}\left(\frac{1}{x^2}\right)\ ,\\
    I^{(2)}_3(u,v,x)&=&\frac{1}{q^2}\left(
        \frac{\sin x}{x}
        - \frac{\sin(u x) \sin(v x)}{u v x^2}\right) \ ,
\end{eqnarray}
\begin{eqnarray}
        I^{(2)}_{i}(u,v,x) &=&  \frac{\cos x}{2q^2uvx} \left\{ -\mathrm{Si}\left(\left(1 - u + v\right)x\right) + \mathrm{Si}\left(\left(1 - u - v\right)x\right)- \mathrm{Si}\left(\left(1 + u - v\right)x\right) + \mathrm{Si}\left(\left(1 + u + v\right)x\right)\right\}\nonumber \\
        &+& \frac{\sin x}{2q^2uvx} \bigg\{ \mathrm{Ci}\left(\left(1 - u + v\right)x\right) + \mathrm{Ci}\left(\left(1 + u - v\right)x\right) - \mathrm{Ci}\left(|1 - u - v|x\right) - \mathrm{Ci}\left(\left(1 + u + v\right)x\right)\nonumber \\
        &+& \ln\left|\frac{1 - (u + v)^2}{1 - (u - v)^2}\right| \bigg\}  \ ,
\end{eqnarray}
where $\mathrm{Si}$ and $\mathrm{Ci}$ functions are defined as follows:
\begin{equation}
    \mathrm{Si}(x)=\int^{x}_0 \mathrm{d}\bar{x}\frac{\sin \bar{x}}{\bar{x}}\ , \quad
    \mathrm{Ci}(x)=-\int^{\infty}_x \mathrm{d}\bar{x}\frac{\cos \bar{x}}{\bar{x}}\ .
\end{equation}
Considering that current observations correspond to the limit $x \to \infty$, taking this limit in the above result and retaining only the leading order yields
\begin{eqnarray}
        I^{(2)}_1&=&
        \frac{27 (u^2 + v^2 - 3)}{4 q^2u^3 v^3 x}
        \bigg(
            \sin x
            \left(
                -4 u v + (u^2 +v^2- 3)
                \ln \left| \frac{3 - (u + v)^2}{3 - (u - v)^2} \right|
            \right)-\pi(u^2 \nonumber\\
        &+& v^2 - 3) \Theta(v + u - \sqrt{3}) \cos x
        \bigg) \ , \label{eq:11I}
\end{eqnarray}
\begin{eqnarray}
I^{(2)}_2&=&
        \frac{\sqrt{3} \left(u^2 - 3 (1-v)^2\right)}{4q^2 u^3 v x}\Bigg[
            \sin x
            \bigg(
                \left(u^2 - 3 (1+v)^2\right)
                \ln \left|
                    \frac{\left(u - \sqrt{3} v\right)^2 - 3}{\left(u + \sqrt{3} v\right)^2 - 3}
                \right| \nonumber\\
         &-&\frac{4 u v \left(u^2 - 9 v^2 + 9\right)}{\sqrt{3} \left(u^2 - 3 (1-v)^2\right)}
            \bigg) + \pi \left(u^2 - 3 (1+v)^2\right) \Theta (u -  \sqrt{3}|1 - v|) \cos x
        \Bigg]\ , \label{eq:22I}
\end{eqnarray}
\begin{eqnarray}
        I^{(2)}_3&=&\frac{1}{q^2}
        \frac{\sin x}{x} \ , \label{eq:33I}\\
        I^{(2)}_i&=&\frac{1}{2q^2uvx}\Bigg[\sin x \times\ln\left|\frac{1 - (u + v)^2}{1 - (u - v)^2}\right| - \cos x \times\pi
        \Bigg] \  , \ (i=4,5,6,7) \ ,\nonumber\\
        \label{eq:ijI}
\end{eqnarray}
where $\Theta(x)$ represents the Heaviside theta function. We have used the following approximations: $\lim_{x\to\pm \infty} \mathrm{Si}(x)=\pm \pi/2$ and $\lim_{x\to \infty} \mathrm{Ci}(x)=0$ in Eq.~(\ref{eq:11I}) and Eq.~(\ref{eq:22I}). The analytic expression given in Eq.~(\ref{eq:11I}) corresponds to the kernel function of second-order \acp{SIGW}, which has been systematically studied and widely cited in previous works \cite{Espinosa:2018eve,Kohri:2018awv,Ananda:2006af,Alabidi:2012ex,Saito:2008jc,Baumann:2007zm}. The kernel functions in Eq.~(\ref{eq:22I})--Eq.~(\ref{eq:ijI}) correspond to the \acp{GW} induced by first-order tensor–scalar and tensor–tensor source terms, and the corresponding results have been systematically investigated in Refs.~\cite{Chang:2022vlv,Picard:2024ekd,Picard:2023sbz,Bari:2023rcw,Picard:2025bwq,Yu:2023lmo}. Recently, Ref.~\cite{Iania:2026mtp} has extended this result to the case of second-order \acp{TSIGW} in the early matter-dominated era.

After obtaining the analytical expression of the kernel function, the expressions of the second-order \acp{GW} in Eq.~(\ref{eq:hf1})--Eq.~(\ref{eq:hfi}) are fully determined. These expressions can subsequently be employed to evaluate the energy density spectrum of the second-order \acp{TSIGW}. More precisely, the total energy density fraction of \acp{GW} up to second order is given by \cite{Maggiore:1999vm}
\begin{equation}\label{eq:Omega}
	\begin{aligned}
		\Omega_{\mathrm{GW},g}^{\mathrm{tot}}(\eta, q)=\Omega_{\mathrm{GW}}^{(1)}(\eta, q)+\Omega_{\mathrm{GW}}^{(2)}(\eta, q)=\frac{1}{6}\left(\frac{q}{a(\eta) H(\eta)}\right)^{2} \mathcal{P}^{\mathrm{tot}}_{h}(\eta, q) \ ,
	\end{aligned}
\end{equation}
where
\begin{eqnarray}\label{eq:P12}
    \mathcal{P}^{\mathrm{tot}}_{h}(\eta, q)=\mathcal{P}^{(1)}_{h}(\eta, q) +\frac{1}{4}\mathcal{P}^{(2)}_{h}(\eta, q) \ .
\end{eqnarray}
In Eq.~(\ref{eq:Omega}), the subscript $g$ indicates that the contribution is generated by Gaussian primordial perturbations. Furthermore, in Eq.~(\ref{eq:P12}), $\mathcal{P}^{(1)}_{h}(\eta, q)$ and $\mathcal{P}^{(2)}_{h}(\eta, q)$ represent the power spectra of \acp{PGW} and second-order \acp{GW}, respectively. The power spectrum of the $n$th-order \acp{GW} can be calculated using the two-point correlation function of the \acp{GW}
\begin{equation}\label{eq:Phn}
  \left\langle h^{\lambda,(n)}_{\mathbf{q}}h^{\lambda^{\prime},(n)}_{\mathbf{q}^{\prime}}\right\rangle=\delta^{\lambda \lambda^{\prime}} \delta\left(\mathbf{q}+\mathbf{q}^{\prime}\right) \frac{2 \pi^2}{q^3} \mathcal{P}_{h}^{(n)}(q,\eta) \ .
\end{equation}
By substituting Eq.~(\ref{eq:Tt1}) together with Eq.~(\ref{eq:hf1})--Eq.~(\ref{eq:hfi}) into Eq.~(\ref{eq:Phn}), we can calculate the total energy density spectrum $\Omega_{\mathrm{GW},g}^{\mathrm{tot}}(q)$ of \acp{GW} in Eq.~(\ref{eq:Omega}). After a series of lengthy but straightforward calculations, we obtain the analytical expression of $\Omega_{\mathrm{GW},g}^{\mathrm{tot}}(q)$ \cite{Kohri:2018awv,Chang:2022vlv,Wu:2025gwt}
\begin{eqnarray}\label{eq:O1+2}
    \Omega_{\mathrm{GW},g}^{\mathrm{tot}}(q)=\Omega_{\mathrm{GW}}^{(1)}(q)+\Omega_{\mathrm{GW}}^{(2)}(q)=\Omega_{\mathrm{GW}}^{(1)}(q)+\Omega_{\mathrm{GW}}^{\phi\phi}(q)+\Omega_{\mathrm{GW}}^{\phi h}(q)+\Omega_{\mathrm{GW}}^{hh}(q) \ ,
\end{eqnarray}
where
\begin{align}
    \Omega_{\mathrm{GW}}^{(1)}&(q)=\frac{x^2}{6}\mathcal{P}^{(1)}_h(\eta,q)= \frac{1}{12}\mathcal{P}_h(q) \ ,
    \label{eq:Oh11}
\end{align}
\begin{align}
    \Omega_{\mathrm{GW}}^{\phi\phi}&(q)= \int_{0}^{\infty} \mathrm{d}v \int_{|1-v|}^{1+v} \mathrm{d}u\,
    \mathcal{P}_{\zeta}(uq) \mathcal{P}_{\zeta}(vq) ~  \frac{3}{1024 u^8 v^8} (u^2 + v^2 - 3)^2 ~ \left[4v^2 - (1 + v^2 - u^2)^2\right]^2 \notag \\
    &\times \left\{ \left[(u^2 + v^2 - 3) \ln \left| \frac{3 - (u + v)^2}{3 - (u - v)^2} \right| - 4uv \right]^2 +  \pi^2 (u^2 + v^2 - 3)^2 \Theta \Big(u + v - \sqrt{3} \Big) \right\} \ ,
    \label{eq:1Oss}
\end{align}
\begin{equation}
    \begin{aligned}
        \Omega_{\mathrm{GW}}^{\phi h}&(q)=\int_{0}^{\infty} \mathrm{d}v \int_{|1-v|}^{1+v} \mathrm{d}u\, \mathcal{P}_{\zeta}(uq) \mathcal{P}_h(vq)   \bigg[ 16v^4 + 24v^2 (1 + v^2 - u^2)^2 + (1 + v^2 - u^2)^4 \bigg] \\
        &  \frac{1}{442368 u^8 v^8}\Bigg\{
        \Bigg[ 4 u v \Big(u^2 - 9(v^2 - 1)\Big)+\sqrt{3}\Big(u^2 - 3(v - 1)^2 \big) \Big(u^2 - 3(v + 1)^2 \Big)\times \\
        &  \ln \left| \frac{3 - (u + \sqrt{3}v)^2}{3 - (u - \sqrt{3}v)^2} \right| \Bigg]^2 + 3\pi^2\Theta\Big(u^2 - 3(v - 1)^2\Big)  \Big(u^2 - 3(v - 1)^2 \Big)^2 \Big(u^2 - 3(v + 1)^2 \Big)^2  \Bigg\} \ , \label{eq:1Ost}
    \end{aligned}
\end{equation}
\begin{equation}
    \begin{aligned}
        \Omega_{\mathrm{GW}}^{hh}&(q)= \int_{0}^{\infty} \mathrm{d}v \int_{|1-v|}^{1+v} \mathrm{d}u
        \mathcal{P}_h(uq) \mathcal{P}_h(vq)  \frac{1}{3145728\, u^8 v^8}
        \Big((u - v)^2 - 1\Big)^2 \Big((u + v)^2 - 1\Big)^2 \\
        &\times \Bigg[
        64 u^2 v^2 \Big(1 + u^4 + v^4 + 6 (u^2 + v^2) + 6 u^2 v^2\Big) + 16 u v \Big(1 + u^6 + v^6 + 15 (u^4 + v^4)  \\
        &+ 15 (u^2 + v^2) + 15 u^2 v^2 (u^2 + v^2 + 6) \Big)
        \ln \left| \frac{1 - (u + v)^2}{1 - (u - v)^2} \right| + \Big(
            1 - 7 u^8 + 4 v^2 + 126 v^4 + 9 v^8 \\
        &+ 116 v^6 - 12 u^6 (5 + 7 v^2)
            + 2 u^4 (7 + 118 v^2 + 35 v^4) + 4 u^2 (13 + 105 v^2 + 151 v^4 + 35 v^6)
        \Big) \\
        & \times \left( \pi^2 + \ln^2 \left| \frac{1 - (u + v)^2}{1 - (u - v)^2} \right| \right)
        \Bigg] \ .
    \end{aligned}
    \label{eq:1Ott}
\end{equation}
In Eq.~(\ref{eq:Oh11})--Eq.~(\ref{eq:1Ott}), we have used the oscillating average relations $\lim_{x\to \infty}\sin^2x=1/2$, $\lim_{x\to \infty}\cos^2x=1/2$, and $\lim_{x\to \infty}\cos{x}\sin{x}=0$ to simplify the expressions of the energy density spectra when $x\to \infty$ \cite{Yuan:2021qgz}. The symbols $\Omega_{\mathrm{GW}}^{\phi\phi}(q)$, $\Omega_{\mathrm{GW}}^{\phi h}(q)$, and $\Omega_{\mathrm{GW}}^{hh}(q)$ denote the energy density spectra induced by the scalar-scalar source, tensor-scalar source, and tensor-tensor source, respectively. Moreover, $\mathcal{P}_{\zeta}(q)$ and $\mathcal{P}_{h}(q)$ represent the primordial power spectra of the curvature perturbation and tensor perturbation, respectively. $\mathcal{P}_{\zeta}(q)$ and $\mathcal{P}_{h}(q)$ can be calculated from the two-point correlation functions of the primordial perturbations in Eq.~(\ref{eq:Ts1}) and Eq.~(\ref{eq:Tt1}), namely:
\begin{equation}\label{eq:Ppzeh}
  \left\langle \zeta_{\mathbf{q}}\zeta_{\mathbf{q}^{\prime}}\right\rangle=\delta\left(\mathbf{q}+\mathbf{q}^{\prime}\right) \frac{2 \pi^2}{q^3} \mathcal{P}_{\zeta}(q) \ , \
  \left\langle h^{\lambda}_{\mathbf{q}}h^{\lambda^{\prime}}_{\mathbf{q}^{\prime}}\right\rangle=\delta^{\lambda \lambda^{\prime}} \delta\left(\mathbf{q}+\mathbf{q}^{\prime}\right) \frac{2 \pi^2}{q^3} \mathcal{P}_{h}(q) \ .
\end{equation}
Given specific primordial power spectra $\mathcal{P}_{\zeta}(q)$ and $\mathcal{P}_{h}(q)$, we can then use Eq.~(\ref{eq:Oh11})--Eq.~(\ref{eq:1Ott}) to calculate the total energy density spectrum $\Omega_{\mathrm{GW}}^{\mathrm{tot}}(q)$ of \acp{GW} during the \ac{RD} era. Here, it should be noted that when we use Eq.~(\ref{eq:hsum})--Eq.~(\ref{eq:hfi}) to calculate the two-point correlation function of the second-order \acp{TSIGW}, we inevitably encounter the cross-correlation functions among $h_1^{\lambda,(2)}$, $h_2^{\lambda,(2)}$, and $h_i^{\lambda,(2)}$ $(i=3\sim7)$: namely $\langle h_1^{\lambda,(2)} h_2^{\lambda',(2)} \rangle$, $\langle h_1^{\lambda,(2)} h_i^{\lambda',(2)} \rangle$, and $\langle h_2^{\lambda,(2)} h_i^{\lambda',(2)} \rangle$. The calculation of these cross-correlation functions requires consideration of the four-point correlation functions of primordial perturbations, specifically $\langle  \zeta_{\mathbf{q}-\mathbf{k}}\zeta_{\mathbf{k}} \zeta_{\mathbf{q}'-\mathbf{k}'}\mathbf{h}^{\lambda_1'}_{\mathbf{k}'}\rangle$, $\langle   \zeta_{\mathbf{q}-\mathbf{k}}\zeta_{\mathbf{k}} \mathbf{h}^{\lambda_1'}_{\mathbf{q}'-\mathbf{k}'}\mathbf{h}^{\lambda_2'}_{\mathbf{k}'} \rangle$, and $\langle \zeta_{\mathbf{q}-\mathbf{k}}\mathbf{h}^{\lambda_1}_{\mathbf{k}}\mathbf{h}^{\lambda_1'}_{\mathbf{q}'-\mathbf{k}'}\mathbf{h}^{\lambda_2'}_{\mathbf{k}'}  \rangle$. In the above calculations, we assume that the two-point correlation function between the primordial perturbations $\zeta_{\mathbf{q}}$ and $\mathbf{h}^{\lambda_1}_{\mathbf{k}}$ satisfies $\langle \zeta_{\mathbf{q}}\mathbf{h}^{\lambda_1}_{\mathbf{k}} \rangle=0$, which directly leads to $\langle h_1^{\lambda,(2)} h_2^{\lambda',(2)} \rangle=\langle h_2^{\lambda,(2)} h_i^{\lambda',(2)} \rangle=0$ for $i=3\sim 7$. For the two-point correlation function $\langle h_1^{\lambda,(2)} h_i^{\lambda',(2)} \rangle$, under the assumption $\langle \zeta_{\mathbf{q}}\mathbf{h}^{\lambda_1}_{\mathbf{k}} \rangle=0$, we can directly demonstrate through explicit calculation that its contribution to the energy density spectrum of \acp{TSIGW} vanishes. More precisely, evaluating the two-point correlation function of the \acp{GW} in Eq.~(\ref{eq:hf1}) and Eq.~(\ref{eq:hfi}) gives:
\begin{equation}\label{eq:X}
	\begin{aligned}
		\langle h_1^{\lambda,(2)} h_i^{\lambda',(2)} \rangle&=\frac{4}{9}\int\frac{\mathrm{d}^3k}{(2\pi)^{3/2}}\int\frac{\mathrm{d}^3k'}{(2\pi)^{3/2}}\varepsilon^{\lambda,lm}\left(\mathbf{q}\right)\varepsilon^{\lambda',lm}\left(\mathbf{q}'\right)k_lk_m \mathbb{P}_{lm,i}^{\lambda_1'\lambda_2'} \\
&\times I^{(2)}_{1}\left( u,v,x \right)I^{(2)}_{i}\left(u',v',x'\right)\langle\zeta_{\mathbf{q}-\mathbf{k}}\zeta_{\mathbf{k}}\mathbf{h}^{\lambda_1'}_{\mathbf{q}'-\mathbf{k}'}\mathbf{h}^{\lambda_2'}_{\mathbf{k}'} \rangle \ .
	\end{aligned}
\end{equation}
Here, the condition $\langle \zeta_{\mathbf{q}}\mathbf{h}^{\lambda_1}_{\mathbf{k}} \rangle=0$ ensures that the Wick's expansion of the four-point correlation function $\langle\zeta_{\mathbf{q}-\mathbf{k}}\zeta_{\mathbf{k}}\mathbf{h}^{\lambda_1'}_{\mathbf{q}'-\mathbf{k}'}\mathbf{h}^{\lambda_2'}_{\mathbf{k}'} \rangle$ contains only the term $\langle\zeta_{\mathbf{q}-\mathbf{k}}\zeta_{\mathbf{k}}\rangle \langle \mathbf{h}^{\lambda_1'}_{\mathbf{q}'-\mathbf{k}'}\mathbf{h}^{\lambda_2'}_{\mathbf{k}'} \rangle$. By simplifying Eq.~(\ref{eq:X}) using Eq.~(\ref{eq:Ppzeh}), we obtain
\begin{equation}\label{eq:X1}
	\begin{aligned}
		\langle h_1^{\lambda,(2)} h_i^{\lambda',(2)} \rangle&=\left(\frac{4}{9}\int\frac{\mathrm{d}^3k}{(2\pi)^{3/2}}\varepsilon^{\lambda,lm}\left(\mathbf{q}\right)k_lk_mI^{(2)}_{1}\left( u,v,x \right) \delta\left(\mathbf{q}\right) \frac{2 \pi^2}{q^3} \mathcal{P}_{\zeta}(q) \right) \\
&\times \left( \int\frac{\mathrm{d}^3k'}{(2\pi)^{3/2}}\varepsilon^{\lambda',lm}\left(\mathbf{q}'\right) \mathbb{P}_{lm,i}^{\lambda_1'\lambda_2'}I^{(2)}_{i}\left(u',v',x'\right)
   \delta^{\lambda_1^{\prime} \lambda_2^{\prime}} \delta\left(\mathbf{q}^{\prime}\right) \frac{2 \pi^2}{q'^3} \mathcal{P}_{h}(q') \right) \ .
	\end{aligned}
\end{equation}
As shown in Eq.~(\ref{eq:X1}), the two-point correlation function $\langle h_1^{\lambda,(2)} h_i^{\lambda',(2)} \rangle$ is decomposed into the product of two independent momentum integrals. This situation corresponds to the disconnected diagrams in the calculation of the two-point correlation function of induced \acp{GW}. To evaluate the first integral in Eq.~(\ref{eq:X1}), we establish a spherical coordinate system with the momentum $\mathbf{q}$ aligned along the $z$-axis. In this case, the contraction of the polarization
tensor $\varepsilon^{\lambda,lm}(\mathbf{q})$ with momenta $k_l$ and $k_m$ can be expressed as \cite{Domenech:2021ztg}
\begin{equation}\label{eq:lambda}
\varepsilon^{\lambda,lm}(\mathbf{q})k_lk_m=\frac{k^2}{\sqrt{2}} \sin ^2 \theta_q \times \begin{cases}\cos 2 \phi_k, & \lambda=+ \\ \sin 2 \phi_k, & \lambda=\times\end{cases} \ ,
\end{equation}
where $\theta$ and $\phi$ denote the polar angle and azimuthal angle, respectively. It should be noted that in the integrand of the first integral in Eq.~(\ref{eq:X1}), only the part given in Eq.~(\ref{eq:lambda}) depends on the azimuthal angle $\phi_k$. The kernel function and the primordial power spectrum depend solely on the relative orientation between the vectors $\mathbf{q}$ and $\mathbf{k}$, and are independent of the azimuthal angle $\phi_k$. Integrating Eq.~(\ref{eq:lambda}) over  $\phi_p$ yields $\int_0^{2\pi} \varepsilon^{\lambda,lm}(\mathbf{q})k_lk_m \mathrm{d}\phi_k =0 $. Then, the two-point correlation function $\langle h_1^{\lambda,(2)} h_i^{\lambda',(2)} \rangle=0$. Therefore, the evaluation of the energy density spectrum of second-order \acp{TSIGW} requires consideration solely of the three contributions specified in Eq.~(\ref{eq:1Oss})--Eq.~(\ref{eq:1Ott}), while the influence of cross-correlation functions can be neglected. This conclusion can be naturally extended to higher-order calculations of \ac{1PR} diagrams associated with \acp{SIGW} \cite{Zhou:2024ncc}.

\section{Primordial non-Gaussianity and induced GWs}\label{sec:3.0}
In Sec.~\ref{sec:2.0}, we summarized the key results for second-order \acp{TSIGW} in the \ac{RD} era with Gaussian primordial perturbations. In this section, we consider  the second-order \acp{TSIGW} in the presence of local-type non-Gaussian primordial curvature perturbations.

\subsection{Primordial non-Gaussianity and SIGWs}\label{sec:3.1}
For the local-type primordial non-Gaussianity, the primordial curvature perturbation with non-Gaussian features can be represented as a local perturbative expansion around the Gaussian primordial curvature perturbation \cite{Luo:1992er,Verde:1999ij,Cai:2018dig,Caporali:2026qhc}
\begin{equation}
    \zeta(\mathbf{x})=\zeta^g(\mathbf{x})
+F_{\rm NL}\left[\left(\zeta^g(\mathbf{x})\right)^2-\langle(\zeta^g(\mathbf{x}))^2\rangle\right] \ ,
\end{equation}
where the symbol  $\zeta^g$ denotes the Gaussian primordial curvature perturbation. In momentum space, the primordial curvature perturbation can be expressed as
\begin{equation}\label{eq:ngzeta}
\zeta_{\mathbf{q}}=\zeta^g_{\mathbf{q}}+F_{\mathrm{NL}}\int\frac{\mathrm{d}^3\mathbf{n}}{(2\pi)^{3/2}} \left( \zeta^g_{\mathbf{q}-\mathbf{n}}\zeta^g_{\mathbf{n}} -\delta\left( \mathbf{q} \right) \frac{2\pi^2}{|\mathbf{n}|^3} \mathcal{P}_{\zeta}(n)  \right) \ ,
\end{equation}
where $F_{\mathrm{NL}}=\frac{3}{5} f_{\mathrm{NL}}$. It should be emphasized that, in calculations of \acp{SIGW}, the second term inside the parentheses on the right-hand side of Eq.~(\ref{eq:ngzeta}) is usually neglected, since it does not affect the evaluation of the energy-density spectrum of \acp{SIGW}. However, as we will see in Sec.~\ref{sec:3.2}, this term must be retained when computing \acp{TSIGW} in the presence of local non-Gaussianity. The non-Gaussian primordial curvature perturbation in Eq.~(\ref{eq:ngzeta}) affects the expressions of the second-order induced \acp{GW} in Eq.~(\ref{eq:hf1}) and Eq.~(\ref{eq:hf2}), thereby influencing the corresponding energy density spectra. Specifically, for the second-order \acp{SIGW} in Eq.~(\ref{eq:hf1}), the calculation of the associated energy density spectrum involves the four-point correlation function of the primordial curvature perturbation: $\langle h_1^{\lambda,(2)} h_1^{\lambda',(2)} \rangle\sim\langle \zeta_{\mathbf{q}-\mathbf{k}}\zeta_{\mathbf{k}}\zeta_{\mathbf{q}'-\mathbf{k}'}\zeta_{\mathbf{k}'} \rangle$. When considering the non-Gaussian primordial curvature perturbation in Eq.~(\ref{eq:ngzeta}), the energy density spectrum of the second-order \acp{SIGW} can be decomposed into several parts proportional to $(f_{\mathrm{NL}})^0$, $(f_{\mathrm{NL}})^2$, and $(f_{\mathrm{NL}})^4$. The contribution proportional to $(f_{\mathrm{NL}})^0$, generated by Gaussian primordial curvature perturbations, has already been given in Eq.~(\ref{eq:1Oss}). The parts proportional to $(f_{\mathrm{NL}})^2$ and $(f_{\mathrm{NL}})^4$ can be written as \cite{Adshead:2021hnm,Li:2025met}
\begin{equation}\label{eq:ngss2}
\begin{aligned}
\Omega_{\mathrm{GW},2}^{\phi\phi}(\eta, q) &= \frac{F_{\text{NL}}^2}{3\pi} \prod_{i=1}^2 \int_0^\infty dt_i \int_{-1}^1 ds_i \, v_i u_i
\Bigg\{
\frac{\pi \overline{J^2(u_1, v_1, x \to \infty)}}{(u_1 v_1 u_2 v_2)^3}
\mathcal{P}_{\zeta}(v_1 v_2 q) \mathcal{P}_{\zeta}(u_1 q) \mathcal{P}_{\zeta}(v_1 u_2 q) \\
&\quad + \int_0^{2\pi} d\varphi_{12} \cos 2\varphi_{12} \overline{J(u_1, v_1, x \to \infty) J(u_2, v_2, x \to \infty)} \\
&\quad\times\frac{\mathcal{P}_{\zeta}(v_2 q) \mathcal{P}_{\zeta}(w_{12} q)}{v_2^3 w_{12}^3}\left[ \frac{\mathcal{P}_{\zeta}(u_1 q)}{u_1^3} + \frac{\mathcal{P}_{\zeta}(u_2 q)}{u_2^3} \right]
\Bigg\} \ ,
\end{aligned}
\end{equation}
\begin{equation}\label{eq:ngss4}
\begin{aligned}
\Omega_{\mathrm{GW},4}^{\phi\phi}(\eta, q) &= \frac{F_{\text{NL}}^4}{24\pi^2} \prod_{i=1}^3 \int_0^\infty dt_i \int_{-1}^1 ds_i \, v_i u_i
\Bigg\{
\frac{2\pi^2 \overline{J^2(u_1, v_1, x \to \infty)}}{(u_1 v_1 u_2 u_3 v_3)^3}
\mathcal{P}_{\zeta}(v_1 v_2 q) \mathcal{P}_{\zeta}(v_1 u_2 q)\mathcal{P}_{\zeta}(u_1u_3 q) \\
&\quad \times\mathcal{P}_{\zeta}(u_1 v_3 q) + \int_0^{2\pi} d\varphi_{12} d\varphi_{23} \cos 2\varphi_{12}
\overline{J(u_1, v_1, x \to \infty) J(u_2, v_2, x \to \infty)} \\
&\quad\times\frac{\mathcal{P}_{\zeta}(u_3 q) \mathcal{P}_{\zeta}(w_{13} q)}{u_3^3 w_{13}^3}\left[
\frac{\mathcal{P}_{\zeta}(v_3 q) \mathcal{P}_{\zeta}(w_{23} q)}{v_3^3 w_{23}^3}
+ \frac{\mathcal{P}_{\zeta}(w_{23} q) \mathcal{P}_{\zeta}(w_{123} q)}{w_{23}^3 w_{123}^3}
\right]
\Bigg\} \ .
\end{aligned}
\end{equation}
In Eq.~(\ref{eq:ngss2}) and Eq.~(\ref{eq:ngss4}), we set $v_i=\frac{q_i}{q}$,  $u_i=\left|\mathbf{q}-\mathbf{q}_i\right|/q$, $s_i = u_i - v_i$, and $t_i = u_i + v_i - 1$. The quantities $y_{i j}$, $w_{123}$, $w_{ij}$, and $J\left(u_i, v_i, x\right)$ are defined respectively as
\begin{equation}
\begin{aligned}
y_{i j}&=\frac{\mathbf{q}_i \cdot \mathbf{q}_j}{q^2}=\frac{\cos \varphi_{i j}}{4} \sqrt{t_i\left(t_i+2\right)\left(1-s_i^2\right) t_j\left(t_j+2\right)\left(1-s_j^2\right)} \\
&\quad+\frac{1}{4}\left[1-s_i\left(t_i+1\right)\right]\left[1-s_j\left(t_j+1\right)\right] \ , \\
w_{123} & =\frac{\left|\mathbf{q}_1+\mathbf{q}_2-\mathbf{q}_3\right|}{q}=\sqrt{v_1^2+v_2^2+v_3^2+2y_{12}-2y_{13}-2y_{23}} \ , \\
w_{i j} & =\frac{\left|\mathbf{q}_i-\mathbf{q}_j\right|}{q}=\sqrt{v_i^2+v_j^2-2y_{i j}} \ ,
\end{aligned}
\end{equation}
\begin{equation}
    J\left(u_i, v_i, x\right)=\frac{x}{8}\left[\left(v_i+u_i\right)^2-1\right]\left[1-\left(v_i-u_i\right)^2\right] I_1^{(2)}\left(u_i, v_i, x\right) \ .
\end{equation}
The overline on $J$ in Eq.~(\ref{eq:ngss2}) and Eq.~(\ref{eq:ngss4}) denotes the oscillation average with respect to $x$.

\subsection{Primordial non-Gaussianity and TSIGWs}\label{sec:3.2}
\begin{figure*}[htbp]
    \captionsetup{
      justification=raggedright,
      singlelinecheck=true
    }
    \centering
	\subfloat[$\langle \zeta^g_{\mathbf{q}-\mathbf{k}-\mathbf{n}} \zeta^g_{\mathbf{n}}  \rangle \langle \zeta^g_{\mathbf{q}'-\mathbf{k}'-\mathbf{n}'} \zeta^g_{\mathbf{n}'} \rangle$ \label{fig:1a}]{\includegraphics[width=.47\columnwidth]{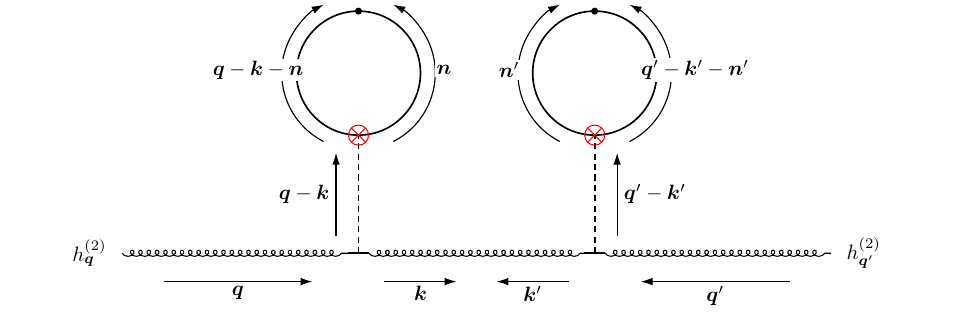}} 
    \hspace{0.1cm}
        \subfloat[$\langle \zeta^g_{\mathbf{q}-\mathbf{k}-\mathbf{n}}  \zeta^g_{\mathbf{q}'-\mathbf{k}'-\mathbf{n}'}  \rangle\langle \zeta^g_{\mathbf{n}} \zeta^g_{\mathbf{n}'} \rangle$\label{fig:1b}] {\includegraphics[width=.47\columnwidth]{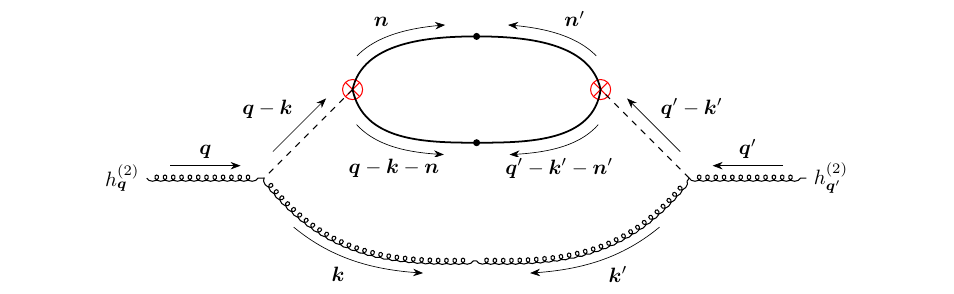}} \\
        \subfloat[$\langle \zeta^g_{\mathbf{q}-\mathbf{k}-\mathbf{n}}  \zeta^g_{\mathbf{n}'} \rangle\langle  \zeta^g_{\mathbf{n}} \zeta^g_{\mathbf{q}'-\mathbf{k}'-\mathbf{n}'}  \rangle$\label{fig:1c}]{\includegraphics[width=.47\columnwidth]{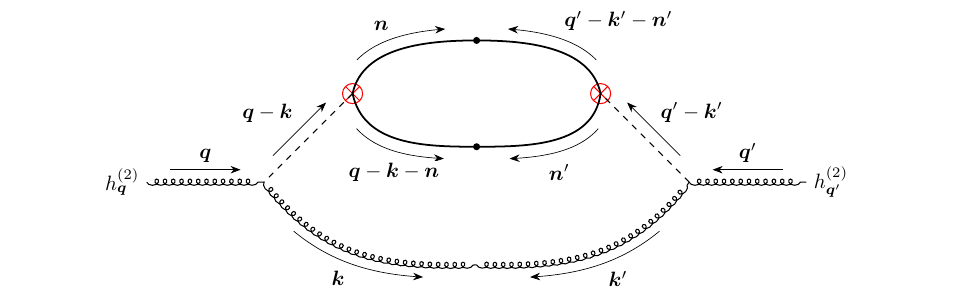}}  
        \hspace{2cm}
        \subfloat[Non-Gaussian vertex.\label{fig:vertex}] {\includegraphics[width=.23\columnwidth]{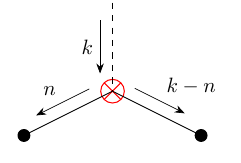}}
\caption{\label{fig:FeynDiag1} The dashed and spring-like lines in the figure represent scalar and tensor perturbations, respectively. Figure (d) depicts the non-Gaussian vertex. Unlike in \acp{SIGW}, the \ac{1PR} diagram (a) in \acp{TSIGW} gives a non‑zero contribution to the total energy density spectrum, and this contribution is cancelled by $\left.
\langle
\zeta_{\mathbf q-\mathbf k}
\zeta_{\mathbf q'-\mathbf k'}
\rangle
\right|_{\rm extra}$ in Appendix.~\ref{Ap:A}.}
\end{figure*}

In Sec.~\ref{sec:3.1}, we reviewed the main results for second-order \acp{SIGW} in the presence of local-type non-Gaussian primordial curvature perturbations. In this subsection, we extend these results to the case of second-order \acp{TSIGW}. By substituting the expression for the second-order gravitational waves induced by the tensor–scalar source term in Eq.~(\ref{eq:hf2}) into Eq.~(\ref{eq:Phn}), we obtain
\begin{equation}\label{eq:fc}
    \begin{aligned}
        \langle h^{\lambda,(2)}_{\mathbf{q}} h^{\lambda',(2)}_{\mathbf{q}'} \rangle&=\frac{4}{9}\int\frac{\mathrm{d}^3k}{(2\pi)^{3/2}} \int\frac{\mathrm{d}^3k'}{(2\pi)^{3/2}} \varepsilon^{\lambda,lm}\left(\mathbf{q}\right) \varepsilon^{\lambda_1}_{lm}\left(\mathbf{k}\right)q^2\varepsilon^{\lambda',ij}\left(\mathbf{q}'\right) \varepsilon^{\lambda'_1}_{ij}\left(\mathbf{k}'\right)q'^2 \\
  &\times I^{(2)}_{2}\left( u,v,x \right) I^{(2)}_{2}\left( u',v',x' \right) \delta^{\lambda_1 \lambda_1^{\prime}} \delta\left(\mathbf{k}+\mathbf{k}^{\prime}\right) \frac{2 \pi^2}{k^3} \mathcal{P}_{h}(k) \\
  &\times\langle \zeta_{\mathbf{q}-\mathbf{k}}  \zeta_{\mathbf{q}'-\mathbf{k}'} \rangle \ .
    \end{aligned}
\end{equation}
The first two lines on the right-hand side are identical to those obtained for second-order \acp{TSIGW} in the Gaussian scenario. The only difference appears in the third line, where the two-point function of the primordial curvature perturbations is treated differently. The explicit evaluation of the two-point function $\langle \zeta_{\mathbf{q}-\mathbf{k}}  \zeta_{\mathbf{q}'-\mathbf{k}'} \rangle$ in Eq.~(\ref{eq:fc}) is presented in Appendix.~\ref{Ap:A}. As indicated by Eq.~(\ref{eq:A1}) in Appendix.~\ref{Ap:A}, local non-Gaussianity gives rise to a four-point correlation function $\langle \zeta^g_{\mathbf{q}-\mathbf{k}-\mathbf{n}} \zeta^g_{\mathbf{n}} \zeta^g_{\mathbf{q}'-\mathbf{k}'-\mathbf{n}'} \zeta^g_{\mathbf{n}'} \rangle$ in the computation of the energy density spectrum of \acp{TSIGW}, and its Wick expansion is provided in Eq.~(\ref{eq:A4}). As illustrated in Fig.~\ref{fig:FeynDiag1}, Fig.~\ref{fig:1a} corresponds to the expression on the second line on the right‑hand side of Eq.~(\ref{eq:A4}), while Fig.~\ref{fig:1b} and Fig.~\ref{fig:1c} correspond to the expressions on the last two lines. As mentioned in Sec.~\ref{sec:2.0}, when evaluating the \ac{1PR} diagrams in the calculations of \acp{SIGW}, the integration over the azimuthal angle $\phi$ causes all \ac{1PR} diagrams to vanish identically, a result that has been systematically established in previous studies \cite{Adshead:2021hnm,Zhou:2024ncc}. However, for the tensor-scalar source contribution to \acp{TSIGW} in the presence of local non-Gaussianity, a direct computation shows that the \ac{1PR} diagram associated with Fig.~1(a) yields a non-zero contribution, and this contribution is precisely canceled by the second term inside the parentheses of the local-type non-Gaussian expression in Eq.~(\ref{eq:ngzeta}). This cancellation is the reason why the second term in Eq.~(\ref{eq:ngzeta}) must be retained when analyzing \ac{TSIGW}. Furthermore, we stress that this cancellation of the \ac{1PR} contribution is a feature unique to scenarios with local non-Gaussianity. In the Gaussian case, the \ac{1PR} diagrams appearing in \acp{TSIGW} generally do not vanish and may contribute non-zero terms \cite{Chen:2022dah}.

Using the results in Appendix.~\ref{Ap:A}, we can simplify the expression for the two-point correlation function in Eq.~(\ref{eq:fc}). More precisely, substituting the result in Eq.~(\ref{eq:A5}) into Eq.~(\ref{eq:fc}) and simplifying, we obtain
\begin{equation}\label{eq:gc}
    \begin{aligned}
 \mathcal{P}^{(2)}_{h}(q)&=\frac{q^3}{9 \pi^2} \frac{1}{(2\pi)^{3/2}}\int\frac{\mathrm{d}^3k}{(2\pi)^{3/2}} \delta_{\lambda \lambda^{\prime}}\varepsilon^{\lambda,lm}\left(\mathbf{q}\right)\varepsilon^{\lambda',ij}\left(\mathbf{q}\right) \delta^{\lambda_1 \lambda_1^{\prime}} \varepsilon^{\lambda'_1}_{ij}\left(\mathbf{k}\right)\varepsilon^{\lambda_1}_{lm}\left(\mathbf{k}\right)q^4 \\
  &\times I^{(2)}_{2}\left( u,v,x \right) I^{(2)}_{2}\left( |q-k|,|k|,\eta \right)  \frac{2 \pi^2}{k^3} \mathcal{P}_{h}(k) \\
  &\times 2 \left(F_{\mathrm{NL}}\right)^2 \int\frac{\mathrm{d}^3\mathbf{n}}{(2\pi)^{3}} \frac{2 \pi^2}{|\mathbf{q}-\mathbf{k}-\mathbf{n}|^3}\frac{2 \pi^2}{|\mathbf{n}|^3} \mathcal{P}_{\zeta}(n)\mathcal{P}_{\zeta}(|\mathbf{q}-\mathbf{k}-\mathbf{n}|)
    \end{aligned}
\end{equation}
By comparing Eq.~(\ref{eq:fc}) with Eq.~(\ref{eq:gc}), we find that, when the non‑Gaussianity in Eq.~(\ref{eq:ngzeta}) is taken into account, the corresponding energy density spectrum of \acp{TSIGW} can be obtained by applying the following replacement to the Gaussian result, namely
\begin{equation}\label{eq:tihuan}
    \begin{aligned}
    &\frac{2 \pi^2}{|\mathbf{q}-\mathbf{k}|^3} \mathcal{P}_{\zeta}(|\mathbf{q}-\mathbf{k}|)  \to 
    &2\left(F_{\mathrm{NL}}\right)^2 \int\frac{\mathrm{d}^3\mathbf{n}}{(2\pi)^{3}} \frac{2 \pi^2}{|\mathbf{q}-\mathbf{k}-\mathbf{n}|^3}\frac{2 \pi^2}{|\mathbf{n}|^3} \mathcal{P}_{\zeta}(n)\mathcal{P}_{\zeta}(|\mathbf{q}-\mathbf{k}-\mathbf{n}|) \ .
    \end{aligned}
\end{equation}
In analogy with the treatment of \acp{SIGW}, we introduce the following variables
\begin{equation}
    \begin{aligned}
    |\mathbf{q}-\mathbf{k}-\mathbf{n}|=u_2|\mathbf{q}-\mathbf{k}| =u_2u|\mathbf{q}| \ \ , \ \  |\mathbf{n}|=v_2|\mathbf{q}-\mathbf{k}|=v_2 u |\mathbf{q}| \ .
    \end{aligned}
\end{equation}
Using the above results, we can simplify Eq.~(\ref{eq:tihuan}) into the following form
\begin{equation}
    \begin{aligned}
    \frac{2 \pi^2}{q^3u^3} \mathcal{P}_{\zeta}(uq) & \to 2\left(F_{\mathrm{NL}}\right)^2 \int\frac{\mathrm{d}^3\mathbf{n}}{(2\pi)^{3}} \frac{2 \pi^2}{(uu_2q)^3}\frac{2 \pi^2}{(v_2uq)^3} \mathcal{P}_{\zeta}(v_2uq)\mathcal{P}_{\zeta}(u_2uq) \\
&=2\left(F_{\mathrm{NL}}\right)^2\int_0^{\infty} \frac{q^3u^3v_2^2\mathrm{d}v_2\mathrm{d}\Omega_n}{(2\pi)^{3}} \frac{2 \pi^2}{(uu_2q)^3}\frac{2 \pi^2}{(v_2uq)^3}~\mathcal{P}_{\zeta}(v_2uq)\mathcal{P}_{\zeta}(u_2uq) \\
&=2\left(F_{\mathrm{NL}}\right)^2\frac{(2\pi^2)^2}{(2\pi)^{3}}\int_0^{\infty} \frac{\mathrm{d}v_2}{v_2}\int_{|1-v_2|}^{1+v_2}\mathrm{d}u_2\frac{u_2}{v_2}\frac{2\pi}{u^3u_2^3q^3} \mathcal{P}_{\zeta}(v_2uq)\mathcal{P}_{\zeta}(u_2uq) \ .
    \end{aligned}
\end{equation}
Then the additional contribution of the local-type non-Gaussian curvature perturbations to the second-order energy density spectrum of \acp{TSIGW} can be written as
\begin{equation}\label{eq:1Ostng}
    \begin{aligned}
        \Omega_{\mathrm{GW},2}^{\phi h}&(q)=\int_{0}^{\infty} \mathrm{d}v \int_{|1-v|}^{1+v} \mathrm{d}u~ \mathcal{P}_h(vq)   \bigg[ 16v^4 + 24v^2 (1 + v^2 - u^2)^2 + (1 + v^2 - u^2)^4 \bigg] \\
        & \times \frac{1}{442368 u^8 v^8}\Bigg\{
        \Bigg[ 4 u v \Big(u^2 - 9(v^2 - 1)\Big)+\sqrt{3}\Big(u^2 - 3(v - 1)^2 \big) \Big(u^2 - 3(v + 1)^2 \Big) \\
        & \times \ln \left| \frac{3 - (u + \sqrt{3}v)^2}{3 - (u - \sqrt{3}v)^2} \right| \Bigg]^2 + 3\pi^2\Theta\Big(u^2 - 3(v - 1)^2\Big)  \Big(u^2 - 3(v - 1)^2 \Big)^2 \Big(u^2 - 3(v + 1)^2 \Big)^2  \Bigg\} \\
&\times \left(F_{\mathrm{NL}}\right)^2\int_0^{\infty} \frac{\mathrm{d}v_2}{v_2^2}\int_{|1-v_2|}^{1+v_2}\frac{\mathrm{d}u_2}{u_2^2} \mathcal{P}_{\zeta}(v_2uq)\mathcal{P}_{\zeta}(u_2uq) \ ,
    \end{aligned}
\end{equation}
where the subscript $"2"$ in $\Omega_{\mathrm{GW},2}^{\phi h}$ in Eq.~(\ref{eq:1Ostng}), similar to Eq.~(\ref{eq:ngss2}) and Eq.~(\ref{eq:ngss4}), indicates that this term is proportional to the square of the non-Gaussian parameter $F_{\mathrm{NL}}$. Combining the above results, the total energy density spectra of  \acp{TSIGW} in the presence of primordial non-Gaussianity can be written as
\begin{equation}\label{eq:totPst}
    \begin{aligned}
        \Omega_{\mathrm{GW}}^{\mathrm{tot}}&(q)=\Omega_{\mathrm{GW},g}^{\mathrm{tot}}(q)+\Omega_{\mathrm{GW},2}^{\phi \phi}(q)+\Omega_{\mathrm{GW},4}^{\phi \phi}(q)+\Omega_{\mathrm{GW},2}^{\phi h}(q) \ ,
    \end{aligned}
\end{equation}
where $\Omega_{\mathrm{GW},g}^{\mathrm{tot}}(q)$ denotes the energy density spectrum of \acp{TSIGW} sourced by Gaussian primordial perturbations, as given in Eq.~(\ref{eq:O1+2}). The terms $\Omega_{\mathrm{GW},2}^{\phi \phi}(q)$ and $\Omega_{\mathrm{GW},4}^{\phi \phi}(q)$ represent the two additional \acp{SIGW} contributions arising from primordial non-Gaussianity, provided in Eq.~(\ref{eq:ngss2}) and Eq.~(\ref{eq:ngss4}). The quantity $\Omega_{\mathrm{GW},2}^{\phi h}(q)$ corresponds to the scalar-tensor source contribution present in the non-Gaussian case, as shown in Eq.~(\ref{eq:1Ostng}). When $F_{\mathrm{NL}}=0$, we have $\Omega_{\mathrm{GW}}^{\mathrm{tot}}(q)=\Omega_{\mathrm{GW},g}^{\mathrm{tot}}(q)$, and the \acp{TSIGW} spectrum reduces to the Gaussian result in Eq.~(\ref{eq:O1+2}). When $\mathcal{P}_h(k)=0$, the small-scale \acp{PGW} are neglected, and the total spectrum becomes $\Omega_{\mathrm{GW}}^{\mathrm{tot}}(q)=\Omega_{\mathrm{GW}}^{\phi\phi}(q)+\Omega_{\mathrm{GW},2}^{\phi \phi}(q)+\Omega_{\mathrm{GW},4}^{\phi \phi}(q)$, consistent with the non-Gaussian second-order \acp{SIGW} results. 

\begin{figure}[t]
    \centering
    \includegraphics[width=0.68\columnwidth]{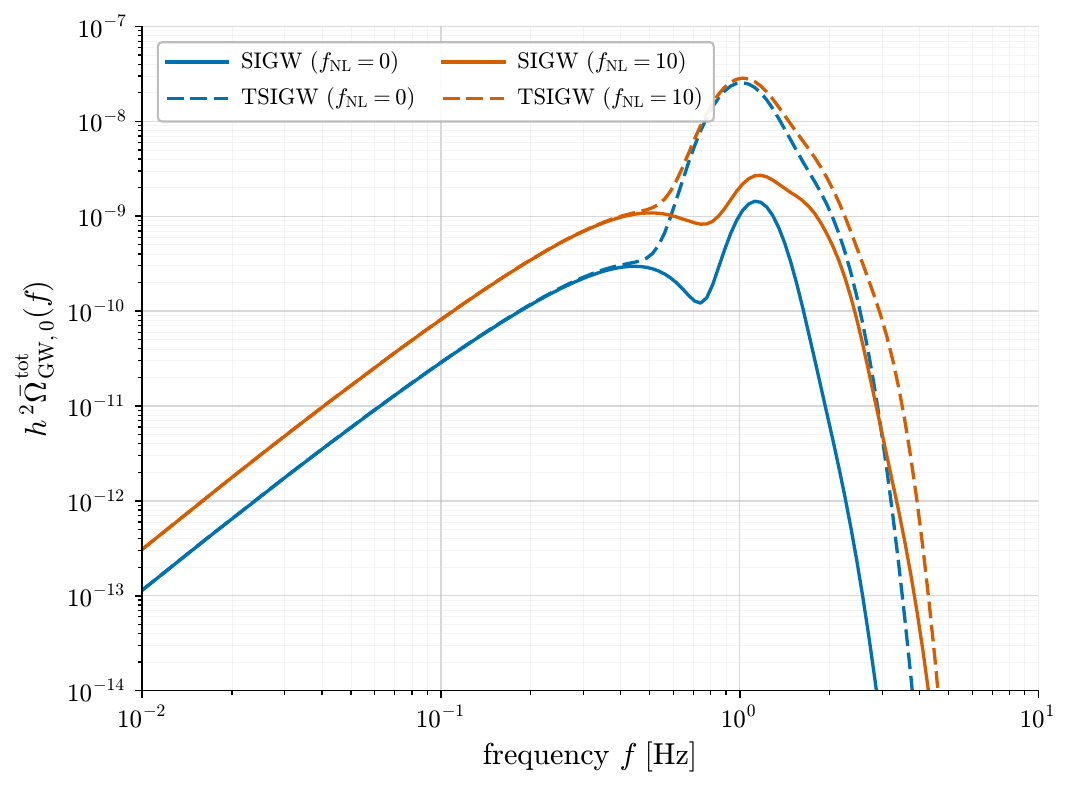}
    \caption{Total energy‑density spectra of second‑order \acp{SIGW} and \acp{TSIGW} obtained for different non‑Gaussian parameter choices. We set $A_{\zeta}=A_h=0.01$, $\sigma=0.2$, and $f_*=1$.}
    \label{fig:sigw&Tsigw-spectrum}
\end{figure}

Furthermore, Eq.~(\ref{eq:totPst}) provides the energy density spectrum of \acp{PGW}$+$\acp{TSIGW} during the \ac{RD} era. Taking into account the thermal history of the universe, the current total energy density spectrum $\bar{\Omega}^{(2)}_{\mathrm{GW,0}}(k)$ is given by
\begin{equation}\label{eq:dangjin}
    \bar{\Omega}^{\mathrm{tot}}_{\mathrm{GW,0}}(k) = \Omega_{\mathrm{rad},0}\left(\frac{g_{*,\rho,e}}{g_{*,\rho,0}}\right)\left(\frac{g_{*,s,0}}{g_{*,s,e}}\right)^{4/3}\bar{\Omega}^{\mathrm{tot}}_{\mathrm{GW}}(k) \ ,
\end{equation}
where $\Omega_{\mathrm{rad},0}$ ($ =4.2\times 10^{-5}h^{-2}$) is the energy density fraction of radiation today. In this paper, we focus on the \ac{LN} primordial power spectra
\begin{equation} \label{eq:LN1}
\mathcal{P}_\zeta (k)=\frac{A_{\zeta}}{\sqrt{2 \pi \sigma^2} } \exp \left(-\frac{1}{2 \sigma^2} \ln ^2\left(k / k_*\right)\right) \ ,
\end{equation}
\begin{equation} \label{eq:LN2}
\mathcal{P}_h (k)=\frac{A_{h}}{\sqrt{2 \pi \sigma^2} } \exp \left(-\frac{1}{2 \sigma^2} \ln ^2\left(k / k_*\right)\right) \ ,
\end{equation}
where $A_{\zeta}$ and $A_h$ are amplitudes of primordial power spectra and $k_*=2\pi f_*$ is the wavenumber at which the primordial power spectrum has a \ac{LN} peak. The parameter $\sigma$ indicates the width of the \ac{LN} primordial power spectrum. Using Eq.~(\ref{eq:totPst}) and Eq.~(\ref{eq:dangjin}), we can calculate the energy density spectra of second-order \acp{SIGW} and second-order \acp{TSIGW} for the \ac{LN} primordial power spectrum. As shown in Fig.~\ref{fig:sigw&Tsigw-spectrum}, we present the contributions from the different components of the total energy density spectrum.

\section{Angular power spectrum of induced gravitational waves}\label{sec:4}
In the previous sections, we discussed the energy density spectrum of second-order induced \acp{GW} in the presence of local-type primordial non-Gaussian curvature perturbations. As mentioned in Sec.~\ref{sec:1.0}, within the \ac{PTA} frequency range, substantial degeneracies may arise both among different \ac{SGWB} models and among distinct parameter choices of a single model. Such degeneracies can severely limit our ability to identify the physical origin of the \ac{PTA} signal. In this section, we investigate the anisotropies of the \ac{SGWB}, which may provide a new observational handle for distinguishing \ac{PTA} sources in future experiments. To evaluate the anisotropies of induced \acp{GW} arising from primordial non-Gaussianity, it is necessary to decompose the primordial curvature perturbations across different scales. More concretely, the primordial curvature perturbation $\zeta^g$ can be decomposed into long-wavelength and short-wavelength modes, denoted $\zeta_L$ and $\zeta_S$ as follows \cite{Li:2023qua}
\begin{equation}\label{eq:S+L}
\zeta^g(\mathbf{q})=\zeta_L(\mathbf{q})+\zeta_S(\mathbf{q})  \ ,
\end{equation}
where the subscripts $S$ and $L$ denote the short-wavelength and long-wavelength modes, respectively. Based on the above decomposition, the power spectrum of primordial curvature perturbations can be separated into the following three components
\begin{equation}\label{eq:SLP}
    \begin{aligned}
        \langle\zeta_{S}(\mathbf{q})\zeta_{S}(\mathbf{q}^{\prime})\rangle & =\delta^{(3)}(\mathbf{q}+\mathbf{q}^{\prime})\frac{2\pi^2}{q^3}P_{S}(q) \ , \\
        \langle\zeta_{L}(\mathbf{q})\zeta_{L}(\mathbf{q}^{\prime})\rangle & =\delta^{(3)}(\mathbf{q}+\mathbf{q}^{\prime})\frac{2\pi^2}{q^3}P_{L}(q) \ , \\
        \langle\zeta_{S}(\mathbf{q})\zeta_{L}(\mathbf{q}^{\prime})\rangle & =0 \ .
    \end{aligned}
\end{equation}
In Eq.~(\ref{eq:SLP}), $P_{L}(q)$ denotes the primordial power spectrum on large scales, which has been tightly constrained by cosmological observations such as the \ac{CMB} and \ac{LSS} to be an approximately scale‑invariant spectrum with an amplitude $A_{L}\approx 2\times 10^{-9}$. The term $P_{S}(q)$ represents the primordial power spectrum on small scales, which is not subject to stringent constraints from current cosmological measurements.

The \acp{SIGW} are generated in the extremely early Universe at very high redshifts, corresponding to a remarkably small cosmological horizon. However, due to the limited angular resolution of current detectors, the signal measured along any line of sight represents an ensemble average over a large number of such horizon patches, making different directions appear nearly identical. If couplings exist between long-wavelength and short-wavelength modes, the energy density of \acp{SIGW} produced by short modes can be modulated by long modes, thereby inducing the anisotropies. This coupling can be induced by local-type primordial non-Gaussianity. More precisely, the initial inhomogeneities of \acp{SIGW} at a spatial position $\mathbf{x}$ can be characterized by the density contrast $\delta_{\rm gw}(\eta,\mathbf{x},q)$, defined as
\begin{equation}
\delta_{\rm gw}(\eta,\mathbf{x},q)
= 4\pi \frac{\omega_{\rm gw}(\eta,\mathbf{x},q)}{\Omega_{\mathrm{GW},0}^{\mathrm{tot}}(q)} - 1 \ .
\label{eq:delta_gw_def}
\end{equation}
Here, the full energy density spectrum $\omega_{\rm gw}(\eta,\mathbf{x},q)$ is defined through the energy density $\rho_{\rm gw}(\eta,\mathbf{x}) 
= \rho_{\rm crit}(\eta) 
\int d^3 q ~ \frac{\omega_{\rm gw}(\eta,\mathbf{x},q)}{q^3}$. In particular, one finds $\omega_{\rm gw} \sim \langle \zeta \zeta \zeta \zeta \rangle_{\mathbf{x}}$, 
where the subscript $\mathbf{x}$ denotes an ensemble average within the horizon 
enclosing $\mathbf{x}$. It should be noted that the long–short mode decomposition given in Eq.~(\ref{eq:S+L}) and Eq.~(\ref{eq:SLP}) applies to Gaussian primordial perturbations. Therefore, when local primordial non-Gaussianity is taken into account, the contribution to $\delta_{\rm gw}$ that is proportional to $\zeta_L$ becomes $\delta_{\rm gw} \sim \zeta_{L} \langle \zeta_{S} \zeta_S \zeta_{S} \zeta_{S} \rangle_{\mathbf{x}}$. It can be seen that, in the presence of local primordial non-Gaussianity, the two-point correlation function associated with  the density contrast $\delta_{\rm gw}(\eta,\mathbf{x},q)$ becomes non-vanishing. The present density contrast, denoted as $\delta_{\rm gw,0}(\mathbf{q})$, can be estimated analytically using the line-of-sight approach. It contains contributions from the primordial inhomogeneities as well as from propagation effects, and is expressed as \cite{Contaldi:2016koz,Bartolo:2019oiq,Bartolo:2019zvb,Li:2023qua}
\begin{equation}
        \delta_{\mathrm{gw},0}(\mathbf{q})=\left\{F_{\mathrm{NL}} \frac{\Omega_{\mathrm{ng}}\left(\eta, q\right)}{\Omega_{\mathrm{GW}}^{\mathrm{tot}}(\eta,q)}+\frac{3}{5}\left[4-n_{\mathrm{gw}}(q)\right]\right\}
        \times\int \frac{\mathrm{d}^{3} \mathbf{k}}{(2 \pi)^{3 / 2}} e^{i \mathbf{k} \cdot \mathbf{x}} \zeta_{L}(\mathbf{k}) \ ,
\end{equation}
where
\begin{equation}
\begin{aligned}
    \Omega_{\mathrm{ng}}(\eta,q)&=2^3 \Omega_{\mathrm{GW}}^{\phi \phi}(q)+2^2\Omega_{\mathrm{GW},2}^{\phi \phi}(q)+2^2\Omega_{\mathrm{GW}}^{\phi h}(q) \ , \\
    n_{\mathrm{gw}}(q)&=\frac{\partial \ln \Omega_{\mathrm{GW},0}^{\mathrm{tot}}(q)}{\partial \ln q} \approx \frac{\partial \ln \Omega_{\mathrm{GW}}^{\mathrm{tot}}(q)}{\partial \ln q} \ .
   \end{aligned}
\end{equation}
The statistical properties are encoded in the corresponding reduced angular power spectrum, defined via the two‑point function of the current density contrast
\begin{equation}
\left\langle 
\delta_{\text{gw,0},\ell m}(q)\,
\delta^{*}_{\text{gw,0},\ell' m'}(q)
\right\rangle=\delta_{\ell\ell'}\delta_{mm'}\,\tilde{C}_{\ell}(f) \ ,
\label{eq:25}
\end{equation}
where $\delta_{\text{gw,0}}(\mathbf{q})$ is expanded in spherical harmonics
\begin{equation}
\delta_{\text{gw,0}}(\mathbf{q})
= \sum_{\ell m} 
\delta_{\text{gw,0},\ell m}(q)
Y_{\ell m}(\mathbf{n}) \ .
\label{eq:26}
\end{equation}
The explicit expression for the reduced angular power spectrum is given by \cite{Li:2023qua}
\begin{equation}
\tilde{C}_{\ell}(f)
= \frac{18\pi\,\mathcal{P}_{L}^{2}}{25\,\ell(\ell+1)}
\left\{
f_{\text{NL}}
\frac{\Omega_{\text{ng}}(\eta,q)}
     {\Omega_{\text{gw}}(\eta,q)}
+ \left[4 - n_{\text{gw,0}}(f)\right]
\right\}^{2} \  .
\label{eq:27}
\end{equation}
The corresponding angular power spectrum can be written as
\begin{equation}
C_{\ell}(f)
= \left[
\frac{\Omega_{\mathrm{GW},0}^{\mathrm{tot}}(f)}{4\pi}
\right]^{2}
\tilde{C}_{\ell}(f) \ .
\label{eq:28}
\end{equation}
With the theoretical framework established above, we are able to calculate the angular power spectrum of induced \acp{GW} for any specified primordial power spectrum. As will be shown in the next section, the angular power spectrum provides a means to partially break the degeneracies among parameters and models.

\begin{table}[t]
\centering
\small
\setlength{\tabcolsep}{5pt}
\renewcommand{\arraystretch}{1.25}
\begin{tabularx}{\linewidth}{@{}L{0.32\linewidth}Y@{}}
\toprule
\textbf{Source of the SGWB}
&
\textbf{Characteristic angular power spectrum}
\\
\midrule

\acp{SIGW} with local-type primordial non-Gaussianity
&
Approximately scale-invariant in multipole space,
$
C_\ell \propto \frac{1}{\ell(\ell+1)} .
$
The frequency dependence is model dependent and is inherited from the SIGW
monopole and the non-Gaussian contribution~\cite{Bartolo:2019zvb,Bartolo:2019oiq,Li:2023qua}.
\\
\midrule

Compact-binary backgrounds, e.g. astrophysical BBHs or PBH binaries
&
On large angular scales, the angular spectrum is typically found to scale as
$C_\ell \propto 1/(\ell+1/2)$,
or approximately $C_\ell \sim \ell^{-1}$~\cite{Cusin:2018rsq,Wang:2021djr,Bellomo:2021mer}.
\\
\midrule

Cosmic string loops
&
Approximately white in multipole space,
$C_\ell \propto \ell^0$,
up to the detailed dependence on the cosmic-string network model and the loop
emission channel~\cite{Jenkins:2018nty,LISACosmologyWorkingGroup:2022jok}.
\\
\midrule

First-order phase-transition backgrounds
&
For scale-invariant primordial modulation, the induced anisotropy can have
$C_\ell^{\rm GW} \propto 1/[\ell(\ell+1)]$.
The frequency dependence is strongly controlled by the phase-transition
parameters~\cite{Geller:2018mwu,Kumar:2021ffi}.
\\
\midrule

Cosmic domain walls
&
The angular spectrum can also be approximately scale-invariant,
$\ell(\ell+1)C_\ell \simeq \mathrm{const.}$.
Its amplitude can be potentially large compared with ordinary cosmological
anisotropies~\cite{Liu:2020mru}.
\\
\bottomrule
\end{tabularx}
\caption{Typical multipole dependence of SGWB angular power spectra for different sources.}
\label{tab:sgwb_angular_spectra}
\end{table}

\section{Observational constraints}\label{sec:Obs}
In the preceding sections, we presented the theoretical results for the energy‑density spectra and angular power spectra of second‑order \acp{SIGW} and \acp{TSIGW} in the presence of local non‑Gaussianity. In this section,  we apply these results to PTA observations and to the analysis of specific models. More precisely, the likelihood is constructed from the \ac{KDE} representations of the free spectra~\cite{Mitridate:2023oar,Lamb:2023jls}
\begin{equation}
    \ln \mathcal{L}(d|\theta)=\sum_{i=1}^{N_f} p(\Phi_i,\theta) \ .
\end{equation}
Here, $p(\Phi_i,\theta)$ denotes the probability of observing $\Phi_i$ given the parameter set $\theta$, with $\Phi_i=\Phi(f_i)$ the corresponding time delay
\begin{equation}
\Phi(f)=\sqrt{\frac{H_0^2\,\Omega_{\mathrm{GW}}(f)}{8\pi^2 f^5 T_{\mathrm{obs}}}} \ ,
\end{equation}
where $H_0=h\times100\,\mathrm{km\,s^{-1}\,Mpc^{-1}}$ is the present Hubble constant. In this work, we employ the \ac{KDE} representation of the first 14 HD-correlated frequency bins from the NANOGrav 15-year free spectrum~\cite{Nanograv:KDE}. Bayesian inference is performed using \textsc{bilby}  with the \textsc{dynesty} nested sampler~\cite{bilby_paper,Speagle:2019ivv,dynesty_software}. Furthermore, to rigorously assess the viability of different scenarios in accounting for the current \ac{PTA} observations, we also examine a mixed case in which both \acp{SMBHB} and induced \acp{GW} contribute to the observed signal. The energy-density spectrum of \acp{SMBHB} is given by~\cite{NANOGrav:2023hvm}
\begin{equation}
    \Omega_{\mathrm{GW}}^{\mathrm{BH}}(f)
    = \frac{2\pi^2 A_{\mathrm{BHB}}^2}{3H_0^2 h^2}
      \left(\frac{f}{\mathrm{year}^{-1}}\right)^{5-\gamma_{\mathrm{BHB}}}
      \mathrm{year}^{-2}\, .
\end{equation}
The prior distribution for $(\log_{10}A_{\mathrm{BHB}},\gamma_{\mathrm{BHB}})$ is modeled as a multivariate normal with mean and covariance
\begin{equation}
\begin{aligned}
    \boldsymbol{\mu}_{\mathrm{BHB}} &= 
    \begin{pmatrix} -15.6 \\ 4.7 \end{pmatrix}, \\
    \boldsymbol{\sigma}_{\mathrm{BHB}} &= 
    0.1 \times
    \begin{pmatrix}
        2.8 & -0.026 \\
        -0.026 & 2.8
    \end{pmatrix}.
\end{aligned}
\end{equation}

\begin{figure}[htbp]
\centering
\subfloat[The posterior distributions of \ac{SIGW}.]{
    \includegraphics[width=0.4\linewidth]{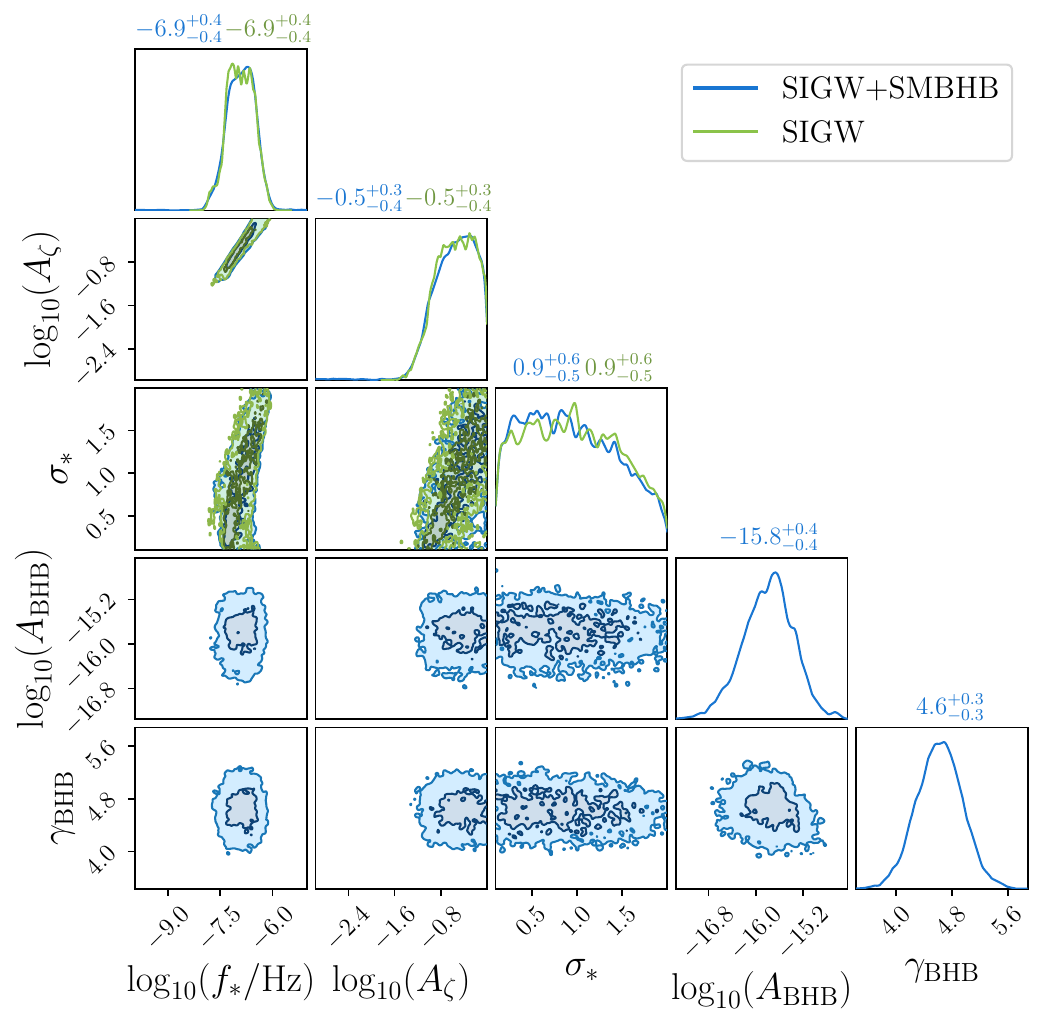}
}\hfill
\subfloat[The posterior distributions of \ac{TSIGW}.]{
    \includegraphics[width=0.4\linewidth]{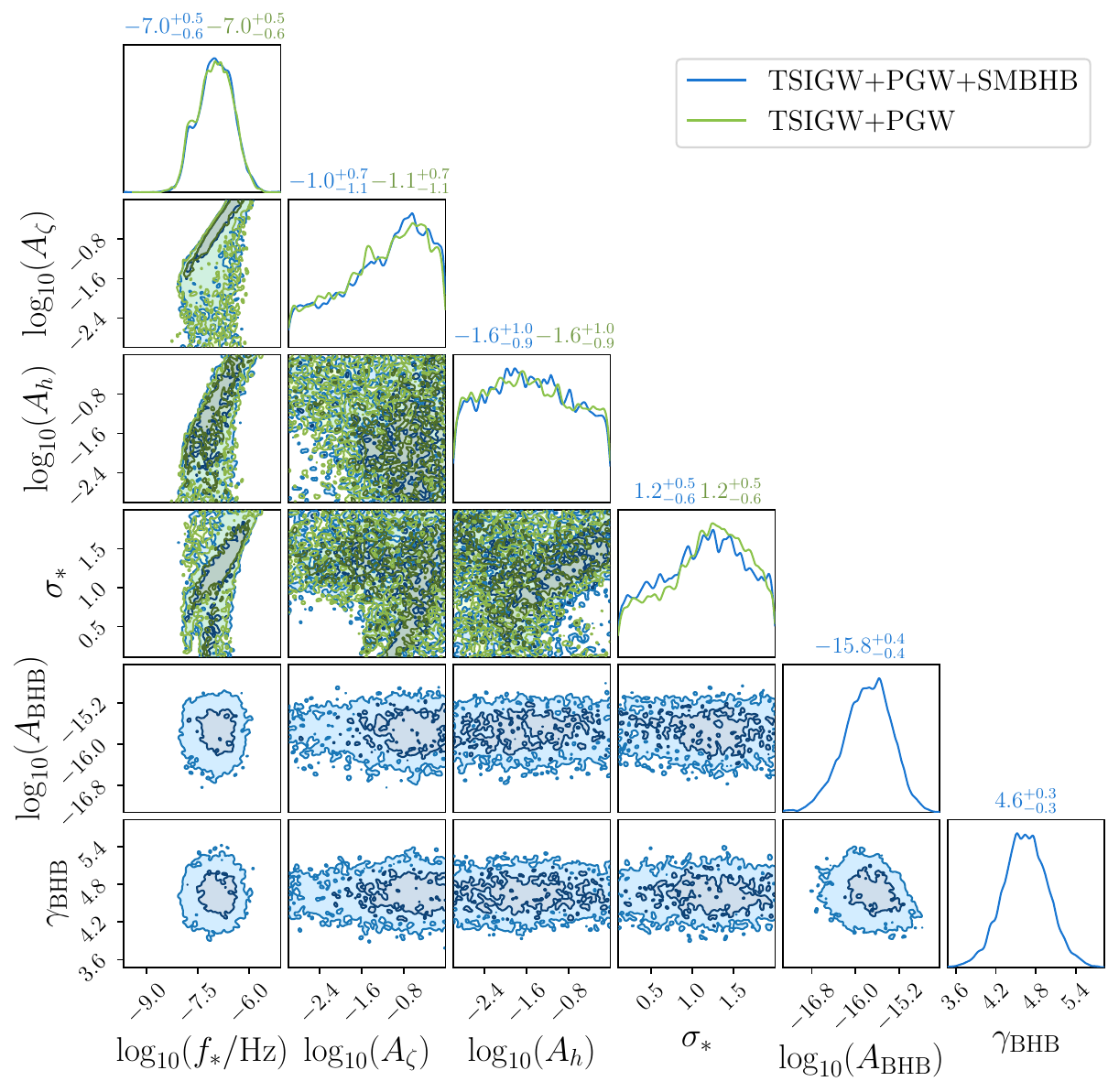}
}
\caption{The posterior corner plots for the \ac{SMBHB} and the induced \acp{GW} considered in the \ac{PTA} Bayesian analysis. The contours in the off-diagonal panels denote the $68\% $ and $95 \%$ credible intervals of the 2D posteriors. The numbers above the figures represent the median values and $1$-$\sigma$ ranges of the parameters. The green and blue curves in the figure correspond to the results obtained from induced \acp{GW} and from the induced \acp{GW} combined with the \ac{SMBHB} contribution, respectively.}
\label{fig:pta_corner_plots1}
\end{figure}

\begin{figure}[htbp]
\centering
\subfloat[The posterior distributions of \ac{SIGW} with local non-Gaussianity.]{
    \includegraphics[width=0.4\linewidth]{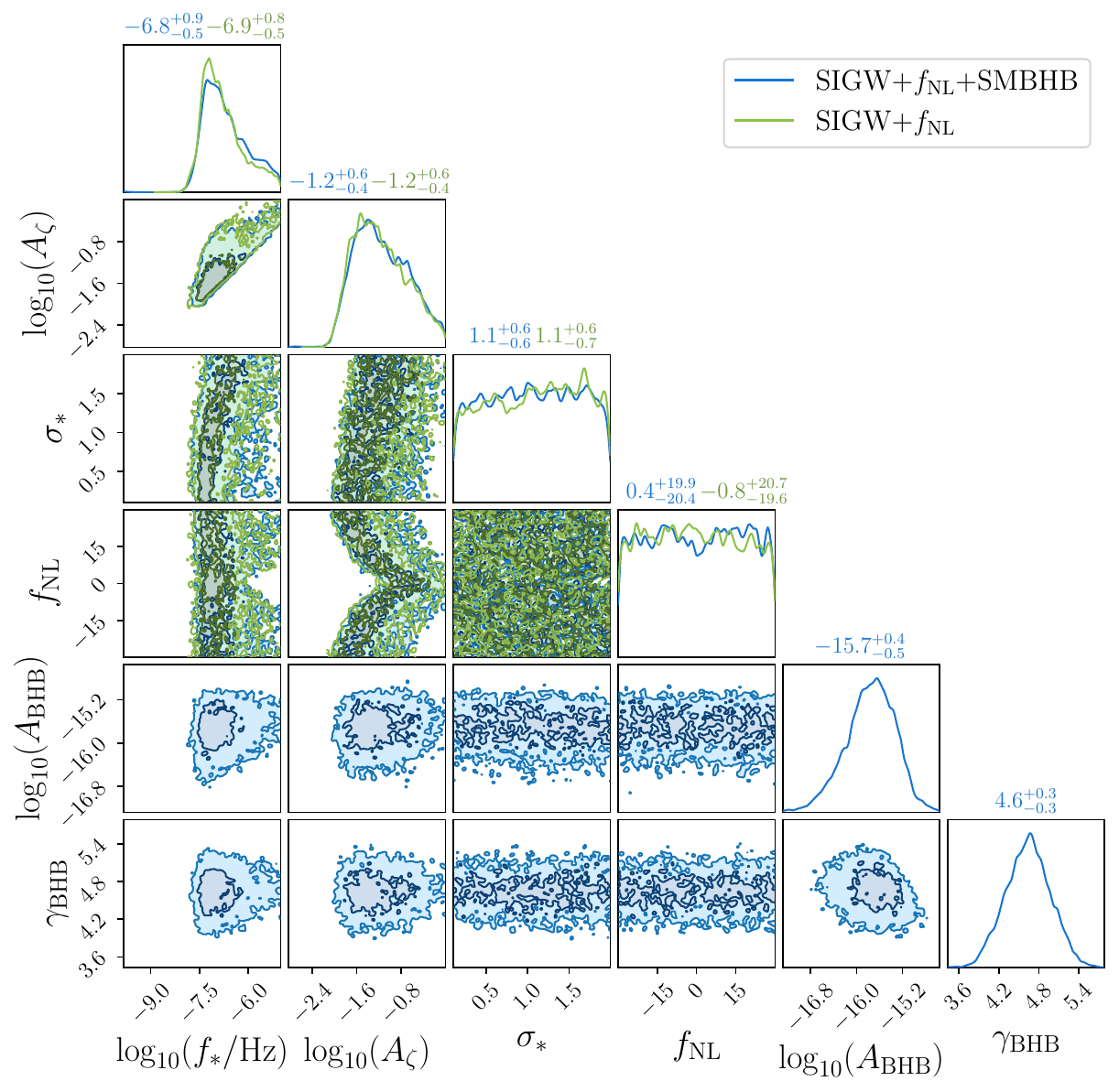}
}\hfill
\subfloat[The posterior distributions of \ac{TSIGW} with local non-Gaussianity.]{
    \includegraphics[width=0.4\linewidth]{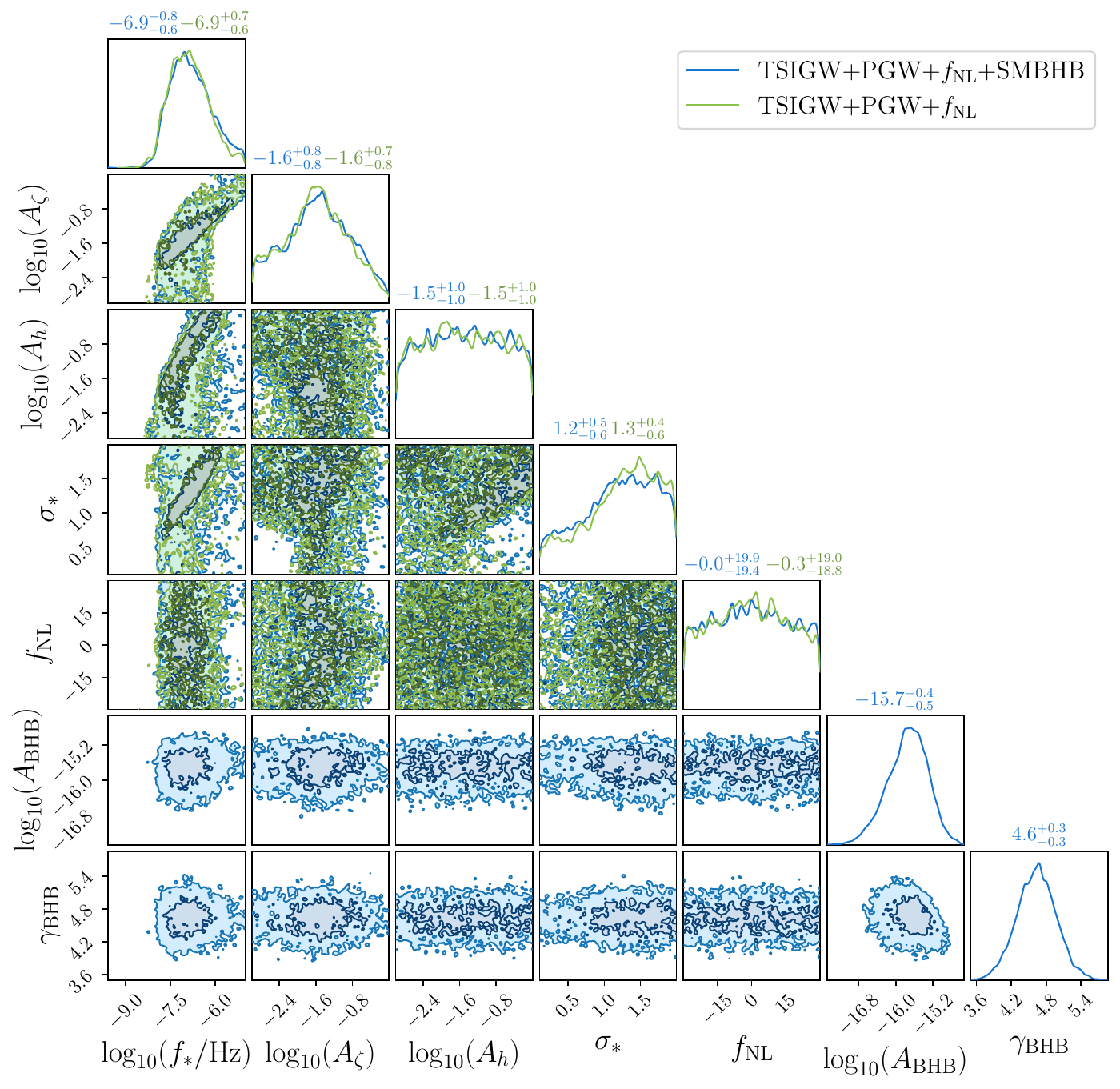}
}
\caption{The green and blue curves in the figure correspond to the results obtained from induced \acp{GW} and from the induced \acp{GW} combined with the \ac{SMBHB} contribution, respectively.}
\label{fig:pta_corner_plots2}
\end{figure}

\begin{figure}[htbp]
\centering
\includegraphics[width=0.68\linewidth]{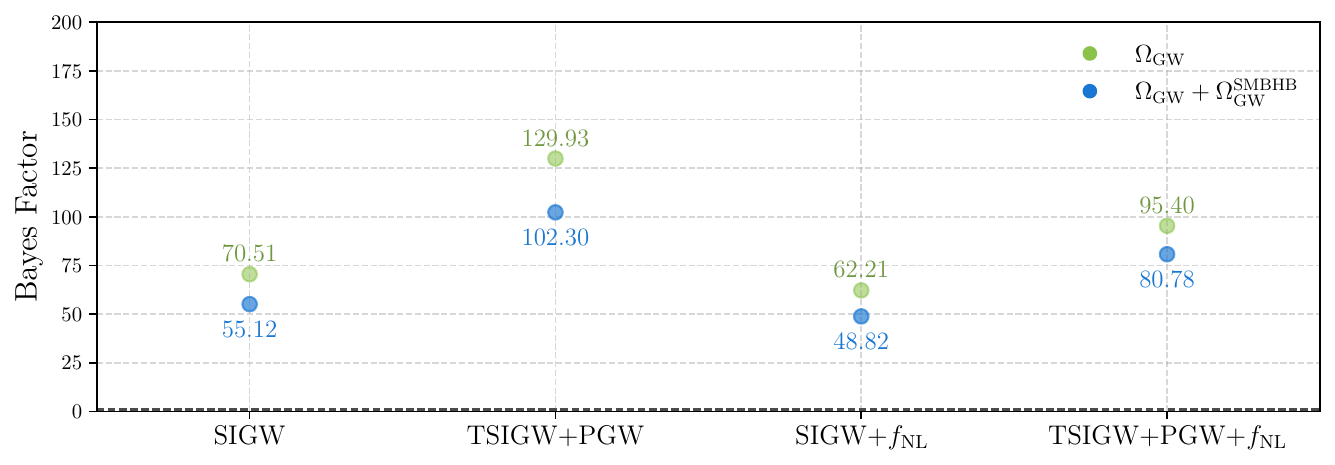}
\caption{Draft Bayes factors for the \ac{SIGW} and \ac{TSIGW} models, with and without local non-Gaussianity, compared with the \ac{SMBHB} model.}
\label{fig:pta_bayes_factor}
\end{figure}

In Fig.~\ref{fig:pta_corner_plots1} and Fig.~\ref{fig:pta_corner_plots2}, we present the posterior distributions corresponding to several different models. It can be seen that, for induced \acp{GW}, current \ac{PTA} observations only place partial constraints on parameters $A_{\zeta}$ and $f_*$. For the remaining parameters of the induced \acp{GW}, the present data do not provide effective constraints. Due to the limited accuracy of current \ac{PTA} measurements and the degeneracies among parameters in induced \acp{GW}, stringent constraints on the parameter space of \acp{SIGW} or \acp{TSIGW} cannot yet be achieved. Furthermore, Fig.~\ref{fig:pta_bayes_factor} displays the Bayes factors for cases in which different models dominate the \ac{PTA} signal. As seen in the figure, although the Bayes factors vary among models, no significant multiplicative differences arise, making it difficult to determine which induced \acp{GW} model is favored by current \ac{PTA} data.

\begin{figure}[htbp]
    \centering
    \includegraphics[width=\columnwidth]{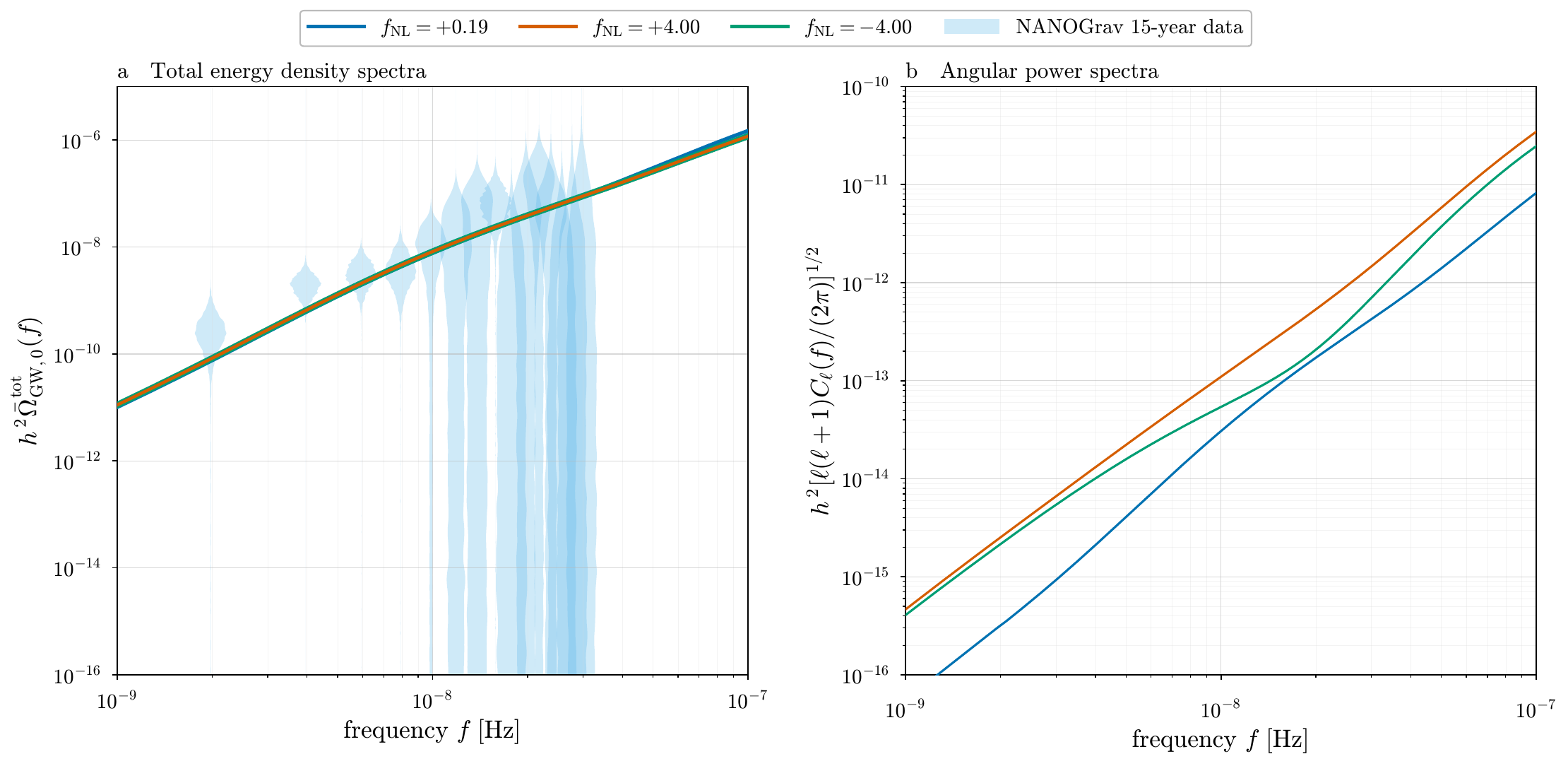}
    \caption{The left panel shows the energy density spectra of \acp{TSIGW}, while the right panel presents the corresponding angular power spectra. 
    All three energy density spectra are evaluated with $\log_{10}(f/\mathrm {Hz})=-6.9$, $\log_{10}A_h=-0.5000$, and $\sigma=1.0$. The blue curve is obtained for $f_{\rm NL}=0.19$ and $\log_{10}A_\zeta=-0.9000$, while the orange and green curves correspond to $f_{\rm NL}=+4.00$ and $-4.00$, respectively, under the parameter choice $\log_{10}A_\zeta=-1.0925$.}
    \label{fig:tsigw-energy-angular}
\end{figure}

\begin{figure}[htbp]
    \centering
    \includegraphics[width=\columnwidth]{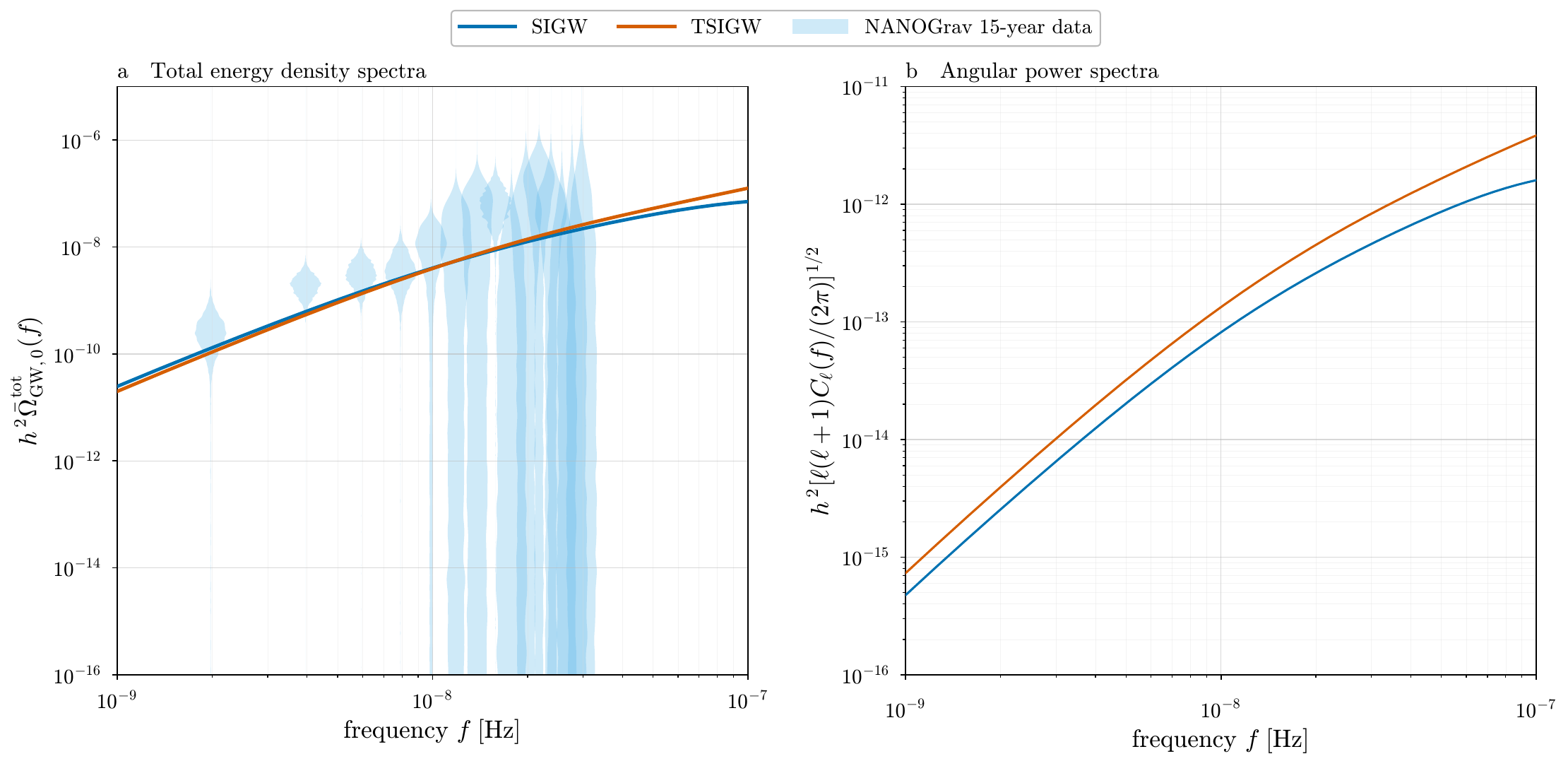}
    \caption{The left panel shows the energy density spectra of \acp{SIGW} and \acp{TSIGW}, while the right panel presents the corresponding angular power spectrum. The SIGW spectrum is computed using $\log_{10}(f/{\rm Hz})=-6.9$, $\log_{10}A_\zeta=-0.718$, $\sigma=1$, $f_{\rm NL}=1$, and $A_h=0$. The TSIGW spectrum is obtained with $\log_{10}(f/{\rm Hz})=-6.9$, $\log_{10}A_\zeta=-0.9$, $\log_{10}A_h=-2$, $\sigma=1$, and $f_{\rm NL}=2.8$.}
    \label{fig:sigw-tsigw-energy-angular}
\end{figure}

To break the degeneracies among different parameters and models, we compute the angular power spectra of \acp{SIGW} and \acp{TSIGW}. As illustrated in Fig.~\ref{fig:tsigw-energy-angular} and Fig.~\ref{fig:sigw-tsigw-energy-angular}, the left panel of Fig.~\ref{fig:tsigw-energy-angular} shows current \ac{PTA} data together with the energy density spectra of \acp{TSIGW}. When \acp{TSIGW} dominate the \ac{PTA} signal, different parameter sets produce almost identical spectra in the \ac{PTA} band, reflecting strong parameter degeneracies. This demonstrates that the energy density spectrum alone is insufficient for disentangling parameter degeneracies within a single model. To address this issue, we compute the corresponding anisotropic angular power spectra, shown in the right panel of Fig.~\ref{fig:tsigw-energy-angular}. Even when the energy density spectra coincide in the \ac{PTA} band for different parameter sets, the corresponding angular power spectra may differ, suggesting that anisotropies of \acp{TSIGW} could serve as an effective means of breaking parameter degeneracies. Moreover, as shown in Fig.~\ref{fig:sigw-tsigw-energy-angular}, we compare the energy density spectra of \acp{SIGW} and \acp{TSIGW} together with their corresponding angular power spectra. We find that degeneracies arise not only among different parameters within the same model, but also across different \ac{SGWB} models, since distinct models may produce very similar energy density spectra.  Nevertheless, even when the spectra coincide, the angular power spectra may differ, providing a potential discriminator between models. Furthermore, beyond induced \acp{GW}, Table.~\ref{tab:sgwb_angular_spectra} summarizes the properties of the angular power spectra for \ac{SGWB} models beyond induced \acp{GW}.

\section{Conclusion and discussion}\label{sec:6}
We systematically investigated the energy density spectrum and anisotropic angular power spectrum of second-order \acp{TSIGW} in the presence of local-type non-Gaussian primordial curvature perturbations. Our results demonstrate that while degeneracies in the energy density spectrum hinder the ability to distinguish \acp{TSIGW} from other \ac{SGWB} sources, the anisotropic angular power spectrum provides a robust diagnostic tool. More precisely, for a given inflationary model that produces large small-scale primordial perturbations, or for a specified parametrization of the primordial power spectrum, the energy density spectrum of induced \acp{GW} depends sensitively on the model or spectral parameters. Varying these theoretical parameters can shift the spectrum’s amplitude, peak frequency, and shape. For a broken power-law spectrum, a \ac{LN} spectrum, or inflationary models featuring parametric resonance, the energy density spectra suffer from strong parameter degeneracy, such that different parameter choices yield very similar energy density outcomes \cite{Zhou:2020kkf,Pi:2020otn,Atal:2021jyo,Atal:2018neu}. This phenomenon does not depend on the current sensitivity of \ac{PTA} observations: even with more accurate \ac{PTA} measurements in the relevant frequency range, the degeneracy among parameters will persist. To address this, we examine the angular power spectrum of the induced \acp{GW}. Our results indicate that parameter choices producing the same energy density spectrum can lead to distinct angular power spectra, so anisotropies may provide a powerful discriminator between different \ac{SGWB} sources and break degeneracies within a given model. In 2023, an upper limit of $\tilde{C_l}<20\%$ was inferred from \ac{PTA} data \cite{NANOGrav:2023tcn}, and future, more precise \ac{PTA} observations could open a new window onto the origin of the \ac{SGWB}.

In addition to analyzing the angular power spectra of \acp{SIGW} and \acp{TSIGW}, we systematically investigated the energy density spectrum of second-order \acp{TSIGW} in the presence of local-type primordial non-Gaussianity. In previous studies of second-order \acp{SIGW}, the case with local-type non-Gaussianity has been systematically analyzed. For \acp{SIGW}, the \ac{1PR} diagrams generated by the Wick expansion vanish exactly after performing the integration over the azimuthal angle $\phi$ in spherical coordinates, implying that \ac{1PR} diagrams do not contribute to the energy density spectrum of \acp{SIGW} \cite{Adshead:2021hnm}. In contrast, in our analysis of second-order \acp{TSIGW}, we find that the two-loop \ac{1PR} diagrams arising in the presence of primordial non-Gaussianity yield a non-zero contribution to the total energy density spectrum of \acp{TSIGW}. Remarkably, this contribution is exactly canceled by the second term $-\delta\left( \mathbf{q} \right) \frac{2\pi^2}{|\mathbf{n}|^3} \mathcal{P}_{\zeta}(n)$ in the parentheses of Eq.~(\ref{eq:ngzeta}). While \acp{SIGW} calculations typically drop this term because it vanishes after $\phi$-integration, the \acp{TSIGW} case requires keeping the full expression, including $-\delta\left( \mathbf{q} \right) \frac{2\pi^2}{|\mathbf{n}|^3} \mathcal{P}_{\zeta}(n)$, to ensure cancellation of the non‑zero \ac{1PR} contribution. Furthermore, it is important to emphasize that in \acp{TSIGW}, the cancellation of the \ac{1PR} diagrams occurs strictly only in the presence of local‑type primordial non-Gaussianity. When the primordial perturbations are Gaussian, the \ac{1PR} diagrams still yield a non‑zero contribution to the total energy density spectrum of \acp{TSIGW} \cite{Chen:2022dah}.

In this work, we focus on \acp{SIGW} and \acp{TSIGW} generated during the standard \ac{RD} era. In this case, the parameters associated with primordial non-Gaussianity and the primordial power spectrum already induce significant degeneracies in the resulting energy density spectra. However, additional physical effects such as a varying sound speed, an early matter-dominated phase, or higher-order cosmological perturbative corrections may arise during the evolution of induced \acp{GW}. These effects further enlarge the model parameter space and can lead to even more severe degeneracies. Current observational probes, including \ac{CMB}, \ac{BAO}, and constraints on $N_{\mathrm{eff}}$, can restrict the parameter space only to a limited extent and are insufficient to rule out most models. Likewise, constraints on \ac{PTA}-band \ac{SGWB} sources from the angular power spectrum will require more precise future observations. To better distinguish different SGWB origins, Ref.~\cite{Wright:2024awr} proposes a general analysis framework, demonstrating that a simple energy conservation argument combined with \ac{CMB} bounds on the radiation energy density can impose stringent limits on model parameters. Future high‑precision \ac{PTA} measurements, together with diversified theoretical approaches, may provide a clearer path toward identifying the sources of the \ac{SGWB}.

\appendix
\section{Two-point function of non-Gaussian primordial curvature perturbations}\label{Ap:A}

As shown in Eq.~(\ref{eq:fc}), when local-type non-Gaussian primordial curvature perturbations are taken into account, the computation of second-order \acp{TSIGW} involves the two-point correlation function $\langle \zeta_{\mathbf{q}-\mathbf{k}}  \zeta_{\mathbf{q}'-\mathbf{k}'} \rangle$ of the primordial curvature perturbations. As shown in Eq.~(\ref{eq:ngzeta}), the primordial curvature perturbation in momentum space can be expressed in the following form
\begin{equation}\label{eq:ngzetaAA}
\zeta_{\mathbf{q}}=\zeta^g_{\mathbf{q}}+F_{\mathrm{NL}}\int\frac{\mathrm{d}^3\mathbf{n}}{(2\pi)^{3/2}} \left( \zeta^g_{\mathbf{q}-\mathbf{n}}\zeta^g_{\mathbf{n}} -\delta\left( \mathbf{q} \right) \frac{2\pi^2}{|\mathbf{n}|^3} \mathcal{P}_{\zeta}(n)  \right) \ .
\end{equation}
By simplifying $\langle \zeta_{\mathbf{q}-\mathbf{k}}  \zeta_{\mathbf{q}'-\mathbf{k}'} \rangle$ using Eq.~(\ref{eq:ngzeta}), we obtain
\begin{equation}\label{eq:A1}
    \begin{aligned}
       \langle \zeta_{\mathbf{q}-\mathbf{k}}  \zeta_{\mathbf{q}'-\mathbf{k}'} \rangle&=\langle \zeta^g_{\mathbf{q}-\mathbf{k}}  \zeta^g_{\mathbf{q}'-\mathbf{k}'} \rangle+\left(F_{\mathrm{NL}}\right)^2 \int\frac{\mathrm{d}^3\mathbf{n}}{(2\pi)^{3/2}}  \int\frac{\mathrm{d}^3\mathbf{n}'}{(2\pi)^{3/2}}  \langle \zeta^g_{\mathbf{q}-\mathbf{k}-\mathbf{n}} \zeta^g_{\mathbf{n}} \zeta^g_{\mathbf{q}'-\mathbf{k}'-\mathbf{n}'} \zeta^g_{\mathbf{n}'} \rangle   \\
       &+\left.\langle
\zeta_{\mathbf q-\mathbf k}
\zeta_{\mathbf q'-\mathbf k'}
\rangle
\right|_{\rm extra} \ ,
    \end{aligned}
\end{equation}
where the symbol $\left.\langle
\zeta_{\mathbf q-\mathbf k}
\zeta_{\mathbf q'-\mathbf k'}
\rangle
\right|_{\rm extra}$ denotes the contribution generated by the second term $\delta\left( \mathbf{q} \right) \frac{2\pi^2}{|\mathbf{n}|^3} \mathcal{P}_{\zeta}(n)$ in Eq.~(\ref{eq:ngzeta}), and its explicit expression is given by
\begin{equation}\label{eq:Aext}
\begin{aligned}
&\left.
\langle
\zeta_{\mathbf q-\mathbf k}
\zeta_{\mathbf q'-\mathbf k'}
\rangle
\right|_{\rm extra}
\\
&=
-\left(F_{\rm NL}\right)^2
\delta(\mathbf q'-\mathbf k')
\int\frac{\mathrm{d}^3\mathbf n'}{(2\pi)^{3/2}}
\frac{2\pi^2}{|\mathbf n'|^3}\mathcal P_\zeta(n')
\int\frac{\mathrm{d}^3\mathbf n}{(2\pi)^{3/2}}
\left\langle
\zeta^g_{\mathbf q-\mathbf k-\mathbf n}\zeta^g_{\mathbf n}
\right\rangle
\\
&\quad
-\left(F_{\rm NL}\right)^2
\delta(\mathbf q-\mathbf k)
\int\frac{\mathrm{d}^3\mathbf n}{(2\pi)^{3/2}}
\frac{2\pi^2}{|\mathbf n|^3}\mathcal P_\zeta(n)
\int\frac{\mathrm{d}^3\mathbf n'}{(2\pi)^{3/2}}
\left\langle
\zeta^g_{\mathbf q'-\mathbf k'-\mathbf n'}\zeta^g_{\mathbf n'}
\right\rangle
\\
&\quad
+\left(F_{\rm NL}\right)^2
\delta(\mathbf q-\mathbf k)\delta(\mathbf q'-\mathbf k')
\int\frac{\mathrm{d}^3\mathbf n}{(2\pi)^{3/2}}
\frac{2\pi^2}{|\mathbf n|^3}\mathcal P_\zeta(n)
\int\frac{\mathrm{d}^3\mathbf n'}{(2\pi)^{3/2}}
\frac{2\pi^2}{|\mathbf n'|^3}\mathcal P_\zeta(n')
\\
&=
-\left(F_{\rm NL}\right)^2
\delta(\mathbf q-\mathbf k)\delta(\mathbf q'-\mathbf k')
\int\frac{\mathrm{d}^3\mathbf n}{(2\pi)^{3/2}}
\frac{2\pi^2}{|\mathbf n|^3}\mathcal P_\zeta(n)
\int\frac{\mathrm{d}^3\mathbf n'}{(2\pi)^{3/2}}
\frac{2\pi^2}{|\mathbf n'|^3}\mathcal P_\zeta(n') \ .
\end{aligned}
\end{equation}
The result obtained in the above equation will play an important role when evaluating one-particle reducible diagrams in second-order \acp{TSIGW}. In addition, the four-point correlation function appearing in Eq.~(\ref{eq:A1}) can be expanded using Wick’s theorem as follows
\begin{equation}\label{eq:A4}
    \begin{aligned}
      &\langle \zeta^g_{\mathbf{q}-\mathbf{k}-\mathbf{n}} \zeta^g_{\mathbf{n}} \zeta^g_{\mathbf{q}'-\mathbf{k}'-\mathbf{n}'} \zeta^g_{\mathbf{n}'} \rangle\\
      &=\langle \zeta^g_{\mathbf{q}-\mathbf{k}-\mathbf{n}} \zeta^g_{\mathbf{n}}  \rangle \langle \zeta^g_{\mathbf{q}'-\mathbf{k}'-\mathbf{n}'} \zeta^g_{\mathbf{n}'} \rangle+\langle \zeta^g_{\mathbf{q}-\mathbf{k}-\mathbf{n}}  \zeta^g_{\mathbf{q}'-\mathbf{k}'-\mathbf{n}'}  \rangle\langle \zeta^g_{\mathbf{n}} \zeta^g_{\mathbf{n}'} \rangle+\langle \zeta^g_{\mathbf{q}-\mathbf{k}-\mathbf{n}}  \zeta^g_{\mathbf{n}'} \rangle\langle  \zeta^g_{\mathbf{n}} \zeta^g_{\mathbf{q}'-\mathbf{k}'-\mathbf{n}'}  \rangle \\
    &=\delta\left(\mathbf{q}-\mathbf{k}\right)\delta\left(\mathbf{q}'-\mathbf{k}'\right) \frac{2 \pi^2}{|\mathbf{n}|^3}\frac{2 \pi^2}{|\mathbf{n}'|^3} \mathcal{P}_{\zeta}(n)\mathcal{P}_{\zeta}(n')\\
    &+\delta\left(\mathbf{q}-\mathbf{k}+\mathbf{q}'-\mathbf{k}'\right)\delta\left(\mathbf{n}'+\mathbf{n}\right) \frac{2 \pi^2}{|\mathbf{q}-\mathbf{k}-\mathbf{n}|^3}\frac{2 \pi^2}{|\mathbf{n}|^3} \mathcal{P}_{\zeta}(n)\mathcal{P}_{\zeta}(|\mathbf{q}-\mathbf{k}-\mathbf{n}|) \\
    &+\delta\left(\mathbf{q}-\mathbf{k}-\mathbf{n}+\mathbf{n}'\right)\delta\left(\mathbf{q}'-\mathbf{k}'-\mathbf{n}'+\mathbf{n}\right) \frac{2 \pi^2}{|\mathbf{q}-\mathbf{k}-\mathbf{n}|^3}\frac{2 \pi^2}{|\mathbf{n}|^3} \mathcal{P}_{\zeta}(n)\mathcal{P}_{\zeta}(|\mathbf{q}-\mathbf{k}-\mathbf{n}|) \ .
    \end{aligned}
\end{equation}
Combining Eq.~(\ref{eq:Aext}) and Eq.~(\ref{eq:A4}), the two-point correlation function $\langle \zeta_{\mathbf{q}-\mathbf{k}}  \zeta_{\mathbf{q}'-\mathbf{k}'} \rangle$ in Eq.~(\ref{eq:A1}) can be simplified to the following form
\begin{equation}\label{eq:A5}
    \begin{aligned}
      \langle \zeta_{\mathbf{q}-\mathbf{k}}  \zeta_{\mathbf{q}'-\mathbf{k}'} \rangle&=\langle \zeta^g_{\mathbf{q}-\mathbf{k}}  \zeta^g_{\mathbf{q}'-\mathbf{k}'} \rangle+2\left(F_{\mathrm{NL}}\right)^2 \delta\left(\mathbf{q}-\mathbf{k}+\mathbf{q}'-\mathbf{k}'\right) \\
 &\times\int\frac{\mathrm{d}^3\mathbf{n}}{(2\pi)^{3}} \frac{2 \pi^2}{|\mathbf{q}-\mathbf{k}-\mathbf{n}|^3}\frac{2 \pi^2}{|\mathbf{n}|^3} \mathcal{P}_{\zeta}(n)\mathcal{P}_{\zeta}(|\mathbf{q}-\mathbf{k}-\mathbf{n}|) \ .
    \end{aligned}
\end{equation}

\acknowledgments
The work is supported in part by the National Natural Science Foundation of China (NSFC) grant  No.12447127.

\bibliography{biblio}

\end{document}